\documentclass[sigconf]{acmart}
\AtBeginDocument{%
  \providecommand\BibTeX{{%
    \normalfont B\kern-0.5em{\scshape i\kern-0.25em b}\kern-0.8em\TeX}}}

\usepackage[ruled]{algorithm2e}
\usepackage{multirow}

\usepackage{enumitem}
\usepackage[most]{tcolorbox}
\usepackage{ifthen}
\usepackage{amssymb}
\usepackage{amsmath}
\usepackage{graphics}
\usepackage{xspace}
\usepackage{balance}

\newboolean{showcomments}
\setboolean{showcomments}{true}
 \setboolean{showcomments}{false}
\ifthenelse{\boolean{showcomments}}
 { \newcommand{\mynote}[2]{
      \fbox{\bfseries\sffamily\scriptsize#1}
        {\small$\blacktriangleright$\textsf{\emph{#2}}$\blacktriangleleft$}}}
        { \newcommand{\mynote}[2]{}}

\newcommand{\toolname}{\textsc{WySiWiM}{}\xspace}

\begin{document}

\title[What You See is What it Means!]{What You See is What it Means! \\ \Large Semantic Representation Learning of Code based on Visualization and Transfer Learning}
\author{Patrick Keller, Laura Plein, Tegawend\'e F. Bissyand\'e, Jacques Klein and Yves Le Traon}
\affiliation{\institution{SnT, University of Luxemburg}}
 \email{firstname.lastname@uni.lu}
\settopmatter{printacmref=false}
\setcopyright{none}



\hyphenation{WYSIWIM}
\begin{abstract}
Recent successes in training word embeddings for NLP tasks have encouraged a wave of research on representation learning for source code, which builds on similar NLP methods. 
The overall objective is then to produce code embeddings that capture the maximum of program semantics.  
State-of-the-art approaches invariably rely on a syntactic representation (i.e., raw lexical tokens, abstract syntax trees, or intermediate representation tokens) to generate embeddings, which are criticized in the literature as non-robust or non-generalizable. 
In this work, we investigate a novel embedding approach based on the intuition that source code has visual patterns of semantics.
We further use these patterns to address the outstanding challenge of identifying semantic code clones. 
We propose the \toolname ({\em ``What You See Is What It Means''}) approach where visual representations of source code are fed into powerful pre-trained image classification neural networks from the field of computer vision to benefit from the practical advantages of transfer learning. 
We evaluate the proposed embedding approach on two variations of the task of semantic code clone identification: code clone detection (a binary classification problem), and code classification (a multi-classification problem). We show with experiments on the BigCloneBench (Java) and Open Judge (C) datasets that although simple, our \toolname approach performs as effectively as state of the art approaches such as ASTNN or TBCNN. We further explore the influence of different steps in our approach, such as the choice of visual representations or the classification algorithm, to eventually discuss the promises and limitations of this research direction.
      
\end{abstract}

\acmYear{2020}


\maketitle

\section{Introduction}
Semantic code clone identification is a long-standing challenge in software engineering~\cite{marcus2001identification}.  It has applications in diverse automation tasks, including bug and vulnerability detection, program repair and synthesis, etc. Until recently, semantic clones were reliably identified using dynamic approaches, such as DyCLINK~\cite{su2016code}, which compare execution traces to decide whether two code fragments behave similarly. Unfortunately, such approaches typically require high-coverage testing to guarantee accuracy. In consequence, they do not effectively scale, and are not usually practical since they require complete and executable code as input. Recent advances in neural networks have provided a new play field for researching static approaches that attempt to learn semantic representations of code via source code embeddings.

Semantic representation learning of source code has attracted significant attention in the research community in the last couple of years~\cite{alon2019code2vec,zhang2019novel,alon2018code2seq,chen2019literature,li2018vuldeepecker,tufano2018deep,wei2017supervised,gao2019teccd,li2017cclearner,krinke_identifying_2001}.
Traditionally, the literature proposes approaches that process code directly or use a syntactic tree representation, where code is treated as sentences. Then, specific approaches inspired by techniques from the Natural Language Processing (NLP) field are used to yield embeddings of these ``sentences''. 
Various works in this realm face robustness issues~\cite{ben2018neural} since simplification of Abstract Syntax Trees (AST), to cope with implementation constraints, weaken the capability of neural network models to capture real and complex semantics~\cite{zhu2015long}.
To address these limitations, the state-of-the-art ASTNN~\cite{zhang2019novel} approach proposes to split each large AST into a sequence of small statement trees, and recursively encodes the statement trees to vectors by capturing the lexical and syntactical knowledge of statements. Although this approach shows promising results on benchmark samples, its reliance on lexical similarity eventually poses two challenges: (1) the model must be regularly trained on new datasets to allow the inner word2vec \cite{mikolov2013efficient} model to capture new vocabulary; (2) the model could be misled by relying on lexical tokens, given that two different library methods with the same names may have different semantics implemented outside the code fragment.


In this paper we propose to investigate another representation learning direction for capturing semantics. In contrast to recent works which all focus on lexical and syntactical information to capture semantics, the intuition behind our approach is to mimic the way a human would instantly perceive code. The data is received through visual perception and forms an image in the head of the programmer. This image is then analyzed for structures the programmer has seen before, by applying his experience. From those recognized structures, the programmer may identify patterns of functionality implementations, which would help him to rapidly infer the semantics of the code fragment, leading to a general understanding of the code. 
We apply the same methodology to design the \toolname (``{\em What You See Is What It Means}'') approach: instead of directly training a complex and opaque semantic representation or embeddings based on syntactical information in source code, we simply render source code into a visual representation that is natural to code. This step is supposed to model the human visual perception. 
After the visualization process, \toolname performs two different procedures. 
First,  the visual structures of the code are extracted.
To that end, a pre-trained image classification neural network, i.e., a neural network that has been trained on other image classification datasets\footnote{The technique of "extracting" knowledge from other datasets is refer to the literature as \emph{transfer learning}} is used to yield a vector of internal features which represent structural information of the input image (i.e., of the code). Optionally, the pre-trained network can be re-trained by adding samples of images representing code.
Second, the feature vectors are used for learning to discriminate between samples implementing different semantics, just as a human developer would do.  Eventually, we expect to leverage the produced classifiers for {\em code classification} (i.e., given a code fragment, predict its functionality label) and {\em clone detection} (i.e., given a pair of code fragments, decide whether they are semantic clones). 

Our main contributions are as follows:
\begin{itemize}
    \item We propose a novel approach to semantic representation learning of code based on visual representations of code fragments. The \toolname approach is intuitively simple, and it builds on transfer learning to efficiently produce embeddings by exploiting powerful pre-trained image classification models from the field of computer vision.
    \item We apply the visual representation embeddings of \toolname to variant tasks of semantic code clone identification. Experimental validations against the BCB and OJ datasets show that \toolname is capable of keeping up with the state-of-the-art while providing significant potential for improvement.
    \item Finally, we provide an analysis of the influence of some implementation choices. Notably, we discuss the possibilities of visual representations of code and the challenges associated to duplicates in the clone benchmarks as part of threats to validity.
\end{itemize}


\section{Background \& Related Work}
We provide in this section an overview of related work after defining essential concepts to facilitate the readers understanding of the \toolname approach description.
\subsection{Definitions}
\subsubsection{Code clone concepts}
We use the following clone-related definitions for our approach in Section~\ref{sec:approach}.

\begin{itemize}[leftmargin=*]
    \item \textbf{Code Fragment: } Also referred to as code snippet, it is a piece of software. Formally, a code fragment is a contiguous set of code lines, which represents the input unit for clone identification. In practice, a code fragment can be a small set of instructions, a whole code block, a whole method or even a whole class.
    \item \textbf{(Code) Clone Pair: } It is a pair of code fragments that are syntactically or semantically similar to each other.
    \item \textbf{Clone Class: } This refers to a set of code fragments where any pairwise combination of code fragments is a clone pair.
    \item \textbf{Syntactic clone pair: } It is a clone pair where the code fragments were deemed similar according to a specific syntactic similarity measure.
    \item \textbf{Semantic clone pair: } It is a clone pair where the code fragments implement the same functionality, respectively the same behaviour or ``semantics''.
    \item \textbf{Candidate pair: }  It refers to a pair of code fragments that may or may not constitute a clone pair. We use this terminology when we do not want to distinguish, or do not yet know, whether or not a pair of code fragments represents a clone pair.
\end{itemize}

We also recall for the reader the following well-accepted definitions of clone types~\cite{bellon2007comparison,roy2009comparison,Su:2016:CodeRelatives}

\begin{itemize}[leftmargin=*]
\item {\bf Type-1}: Identical code fragments, except for differences in whitespace, layout, and comments.
\item {\bf Type-2}: Identical code fragments, except for differences in identifier names and literal values, in addition to Type-1 clone differences. They are also called {\em parameterized} or {\em renamed} clones.
\item {\bf Type-3}: Syntactically similar code fragments that differ at the statement level. The fragments have statements added, modified and/or removed with respect to each other, in addition to Type-1 and Type-2 clone differences. They are also called {\em gapped} or {\em near-miss} clones.
\item {\bf Type-4}: Syntactically dissimilar code fragments that implement the same functionality. They are also known as {\em functional} or {\em semantic} clones. In practice, Type-4 clones are often identified as Type-3 clones with an upper-bound threshold on the syntactic similarity with respect to a specific similarity measure. It should be noted that Type-1 to Type-4 clones are generally considered mutually exclusive.
\end{itemize}
$\Rightarrow$Semantic/functional clones are the primary target in this work. 

\subsubsection{Machine learning concepts}
Since Section~\ref{sec:approach} develops a machine learning approach, we recall for the reader important concepts that we leverage. However, the inner details of these concepts are strictly out of the scope of this work.

\noindent
{\bf Image Classification: } Image classification is a well-studied problem in computer vision with several applications such as facial recognition in smart houses, object recognition for self-driving cars, or disease diagnostic in healthcare. The typical task consists in training a model to classify an image into a single or multiple predefined categories~\cite{kamavisdar2013survey, lu2007survey}. Recent advances in deep neural networks have led to significant breakthroughs in image classification, where computers manage to match human-level accuracy under some conditions~\cite{he2016deep}. Convolutional Neural Networks (CNNs) is the most popular neural network model being used to address the image classification problem. The general idea behind CNNs is that a local understanding of an image is good enough. A convolution is then a weighted sum of the pixel values of the image, as a sliding window is moved across the whole image. Eventually, the CNNs extract low, middle and high-level features and classifiers in an end-to-end multi-layer fashion, and the number of stacked layers can enrich the ``levels''  of features. Simply explained, those image classification neural networks learn to recognize visual features from the images, such as structures and colorings. 

However, CNNs have been shown to present a degradation problem when the deeper network starts to converge: with the network depth increasing, accuracy gets saturated and then degrades rapidly. Residual networks~\cite{he2016deep} (ResNets) have then been proposed to overcome this problem by explicitly letting deeper stacked layers to fit a residual mapping (instead of an underlying mapping as in CNNs). In this work, we will build on these tried and true models from the literature.

\noindent
{\bf Transfer Learning: } Transfer learning is a technique in machine learning which consists of transferring knowledge from a specific domain to another one. To give a real-world example to the concept, we could imagine that learning to play the piano can help a human to learn to play guitar later on. Even though the instruments are very different, the notes and the rhythms are the same, hence we can transfer this knowledge from one task to the other and thus reduce the effort to learn.


\subsection{Related work}
\label{subsec:relatedwork}
Our work is related to various research directions in the literature, including code clone detection, computer vision, machine learning and software engineering benchmarking.

\subsubsection{Code clone identification}
Although code clone identification has been largely studied in the literature, relatively few
techniques have explicitly targeted semantically similar code fragments. Most approaches indeed focus on
textually, structurally or syntactically similar code fragments.
The state-of-the-art techniques on static detection of code clones leverage various intermediate representations to compute code similarity.
Token-based~\cite{baker1993program,kamiya2002ccfinder,li2004cp} representations are used in approaches that target syntactic similarity.
AST-based~\cite{jiang2007deckard,baxter1998clone} representations are employed in approaches that detect similar but potentially
structurally different code fragments. Finally, (program dependency) graph-based~\cite{krinke_identifying_2001,liu_gplag:_2006}
representations are used in detecting clones where statements may be intertwined with each other. Although similar code fragments identified by all these approaches
usually have similar behavior, such static approaches still miss finding such fragments which
have similar behavior even if their code is dissimilar~\cite{juergens2010code}.

To find similarly behaving code fragments, researchers have relied upon dynamic or concolic code similarity detection
which consists in identifying programs that yield similar outputs for the same inputs~\cite{jiang2009automatic,li_measuring_2016,krutz_cccd:_2013,kim-icse2011,Su:2016:CodeRelatives}. Although these
approaches can be very effective in finding semantic code clones, dynamic execution of code is not scalable and implies
several limitations for practical usage (e.g., the need of exhaustive test cases to ensure confidence in behavioral equivalence).

Conceptually, the closest related work is by Ragkhitwetsagul et al.~\cite{ragkhitwetsagul2018picture} who developed a syntactic code clone detection approach based on a visual representation of code. In order to visually represent the code, they pre-process the code by removing comments and normalizing the code formatting. The code is then rendered while applying a syntax highlighting, as done in an IDE, to create the code image. This image is then post-processed by applying various simple image transformations, such as blurring, to finally measure the resulting image similarity. The decision of whether or not two code snippets are clones is then based on the level of image similarity between the visual layout. The scope of such an approach is only limited to syntactic clones. Nevertheless, their experiments also show that the visual representation-based method can generally keep up with the state-of-the-art for syntactic clone detection. 
Although our approach shares the core concept of {\em visualizing code}, we have a  different scope (semantic clones in our case) and we additionally augment this visual representation through transfer learning.

Recently, researchers have investigated leveraging advanced natural language processing and deep learning techniques to statically detect harder-to-detect clones (i.e., type-4 clones). Kim et al.~\cite{facoy} proposed the FaCoY code-to-code search engine where tokens from input code fragments are alternated by considering code fragments from related stackoverflow posts.  This enables the search engine to identify syntactically dissimilar code fragments from the search database. This work, however, is rather competitive to online code search engines than code clone detectors. With their Oreo framework, Saini et al.~\cite{saini2018oreo} have proposed to use a combination of machine learning, information
retrieval, and software metrics to deal with all clone types. They build a specific deep neural network with siamese architecture to address type-4 clones with relative success.

\subsubsection{Semantic representation learning}
Deep learning advances have been exploited for statically learning semantic representations of code. A prominent work in this direction is the Tree-based convolutional neural network (TBCNN) proposed by Mou et al.~\cite{mou2016convolutional}. The authors proposed an effective embedding method for programming language processing, and introduced a large dataset of functional clones which is necessary to train and evaluate the task of code classification.
More recently, Zhang et al.~\cite{zhang2019novel} set the new state-of-the-art representation learning approach with ASTNN, which was demonstrated to be more effective than TBCNN for code clone identification tasks. ASTNN is a semantic embedding method which splits a given code AST into a sequence of smaller statement subtrees and applies a word2vec~\cite{mikolov2013efficient,mikolov2013distributed} embedding to those subtrees. This way ASTNN manages to capture both the lexical and syntactical information within code fragments. We consider both ASTNN and TBCNN as the state-of-the-art for semantic clone identification, and thus they will be used as references for benchmarking our \toolname approach.

\subsubsection{ResNets} Deep Residual Networks~\cite{he2016deep} is undoubtedly today one of the most regarded state-of-the-art techniques within the field of computer vision. This neural network architecture allows to create deeper neural networks for image classification while reducing the network complexity in comparison to other deep learning techniques. Experimental data confirmed that the strategy is effective and may lead to human-level accuracy for the task of image classification. Our approach builds on the success of these networks.

\subsubsection{Benchmarks} 
\label{subsubsec:benchmarks}
In the code clone identification literature, two main benchmarks are widely used.
\begin{itemize}[leftmargin=*]
    \item BigCloneBench (BCB), released by Svajlenko et al.~\cite{svajlenko2014towards}, is the first big-data-curated benchmark of real clones and used to evaluate modern
tools of detecting code clones. It contains 8 million clone pairs and is to the best of our knowledge the biggest publicly available Java code clone benchmarks. It was built by labeling pairs of code fragments from the IJaDataset-2.0~\cite{ijadata}. 
BigCloneBench maintainers have mined this dataset focusing on a specific set of functionalities.

\item OpenJudge (OJ), released by Mou et al.~\cite{mou2016convolutional}, is another public
dataset used to evaluate code-clone detection. It is mostly used in the literature for evaluating program classification approaches, although recent works~\cite{zhang2019novel,saini2018oreo} have applied code clone detection approaches to it. The dataset consists of solutions submitted by students to 104
programming questions on OpenJudge\footnote{\url{http://poj.openjudge.cn/}}, written in C. For each question, there are 500 corresponding solutions, each of
which is verified to be correct by OpenJudge and are thus considered as clones.
\end{itemize}

\section{WYSIWIM}
In this section, we will overview the design of \toolname, providing details on the considered visualizations and the learning models. 
\label{sec:approach}

\subsection{Approach overview}
\label{subsec:overview}
\begin{figure}[!h]
    \centering
    \includegraphics[width=\linewidth]{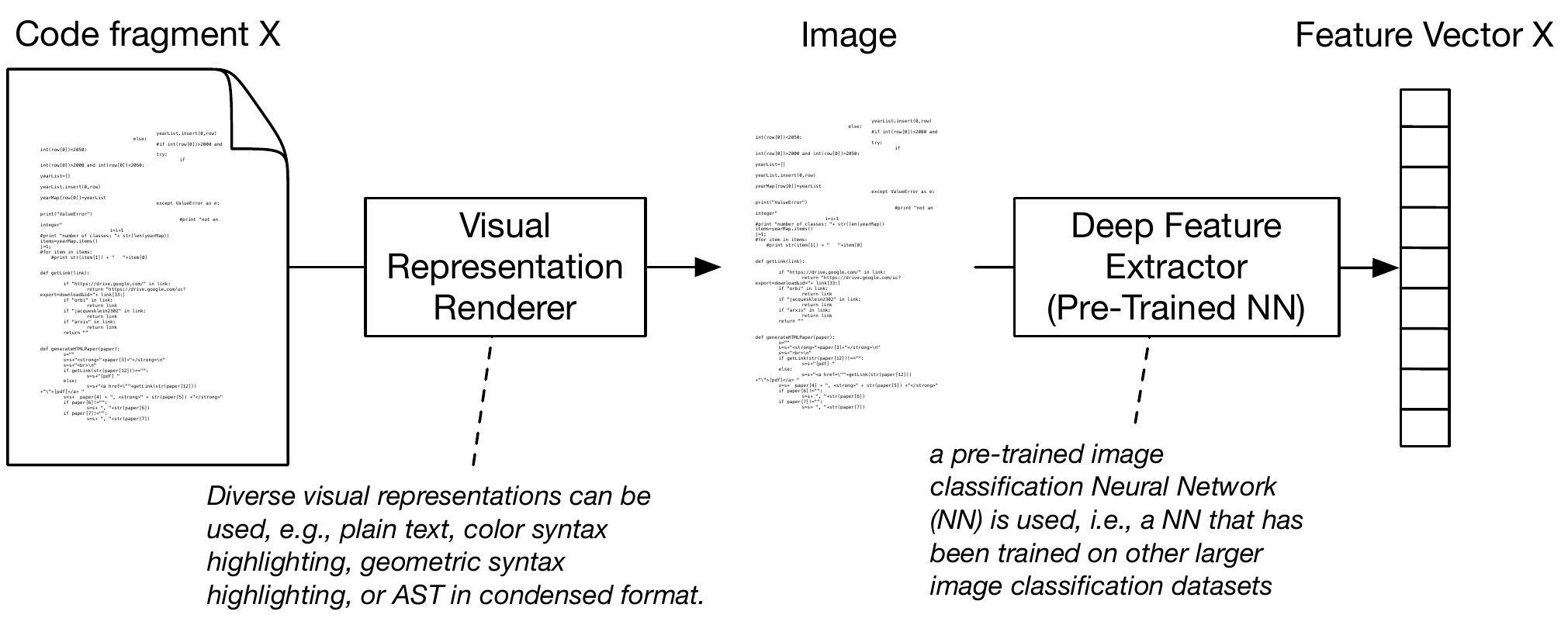}
    \caption{Deep feature extraction (a.k.a, visualization-based code embedding)}
    \label{fig:extractor}
\end{figure}

The core of the \toolname approach is about the production of embeddings for a given code fragment. The idea is to take a code fragment and produce a vector of real numbers so that we receive an actionable representation of the embedded semantic information. As illustrated in Figure~\ref{fig:extractor}, we consider a deep feature extractor which works by producing embeddings for image renderings of code fragments.  

Building a deep feature extractor requires a training step based on a large dataset of images. During such a training, the neural networks learn suitable representations for the images within a feature space. Given that deep neural network architectures for image classification are known to capture a large number of structural features of images, we postulate that pre-trained models can be explored in a transfer learning scenario (cf. Section~\ref{subsec:transfer}).
Transferring the knowledge, embedded in those pre-trained models, allows us to extract visual features without the need of huge amounts of task-specific data to train the feature extractor. 

Once the feature extractor is obtained, one can feed code rendered as images into it to collect the resulting feature vectors. Those can further be used to train simple binary classifiers that learn to apply the embedded semantic information.
Simply put, the deep neural network is used to preprocess images so that they can be used to learn semantics by applying well-known classical machine learning algorithms.

\subsubsection*{Clone identification tasks}
\label{subsec:tasks}
In this work we apply the \toolname approach of visualization-based code semantics learning to the problem of clone identification, which is approached in two different ways:  
as a classical {\em code clone detection} problem and as a {\em code classification} problem.

\noindent
$\star$
In code classification, the goal is to predict the functionality implemented by a code fragment. In practice, we must learn to map the code fragment to one of a set of predefined semantic functionality labels (i.e. {\em clone classes}). It is thus a multi-class classification problem that takes a single code fragment as input and outputs a functionality label.

\noindent
$\star$
In clone detection, the goal is to directly decide if two code fragments are clones. It is thus a binary classification problem that takes a pair of code fragments as input and outputs a Yes/No label on whether or not those fragments form a clone pair.

In principle, both tasks can be emulated by one another. On the one hand, the code classification task could be emulated by finding all clone pairs and building their transitive closure to generate the semantic clone classes. On the other hand, the clone detection task, could be emulated by directly comparing the code fragment labels. We have nevertheless opted in this work to build two separate workflows, both starting by first converting code fragments into their visual representations. 

\noindent
$\star$
For code classification, the collected code ``images'' and their associated functionality labels are used to {\bf fine-tune a pre-trained image classification network}. To that end, the \textbf{size of the output layer} of the pre-trained image classification network must be updated. Indeed the output layer nodes map to the classes that are seen during training. With new datasets, new classes appear.

\noindent
$\star$
For clone detection, the collected code ``images'' are {\bf directly fed into a pre-trained image classification network in order to retrieve the corresponding embeddings} (which are numerical vectors representing the internal structural features within images). Obtained feature vectors are then used for training and testing a classical binary classifier.

\subsection{Transfer learning from pre-trained models}
\label{subsec:transfer}
In our approach, we transfer the embedded knowledge of the pre-trained image classification neural networks to our clone identification tasks (i.e., for both code classification and clone detection).
The knowledge that is transferred in our case is {\em the ability to recognize visual patterns and structures from  images}. Even though the data that those networks are trained on belong technically to a different domain, we expect that they still capture relevant structural knowledge that can be reused to extract the structural information from our specialized (code visualization) images.
Thus, our hypothesis here is that, through the transfer learning, we can leverage powerful pre-trained networks which are able to effectively embed meaningful syntactic as well as semantic structures~\cite{he2016deep}. 

\begin{figure}[!h]
    \centering
    \includegraphics[width=\linewidth]{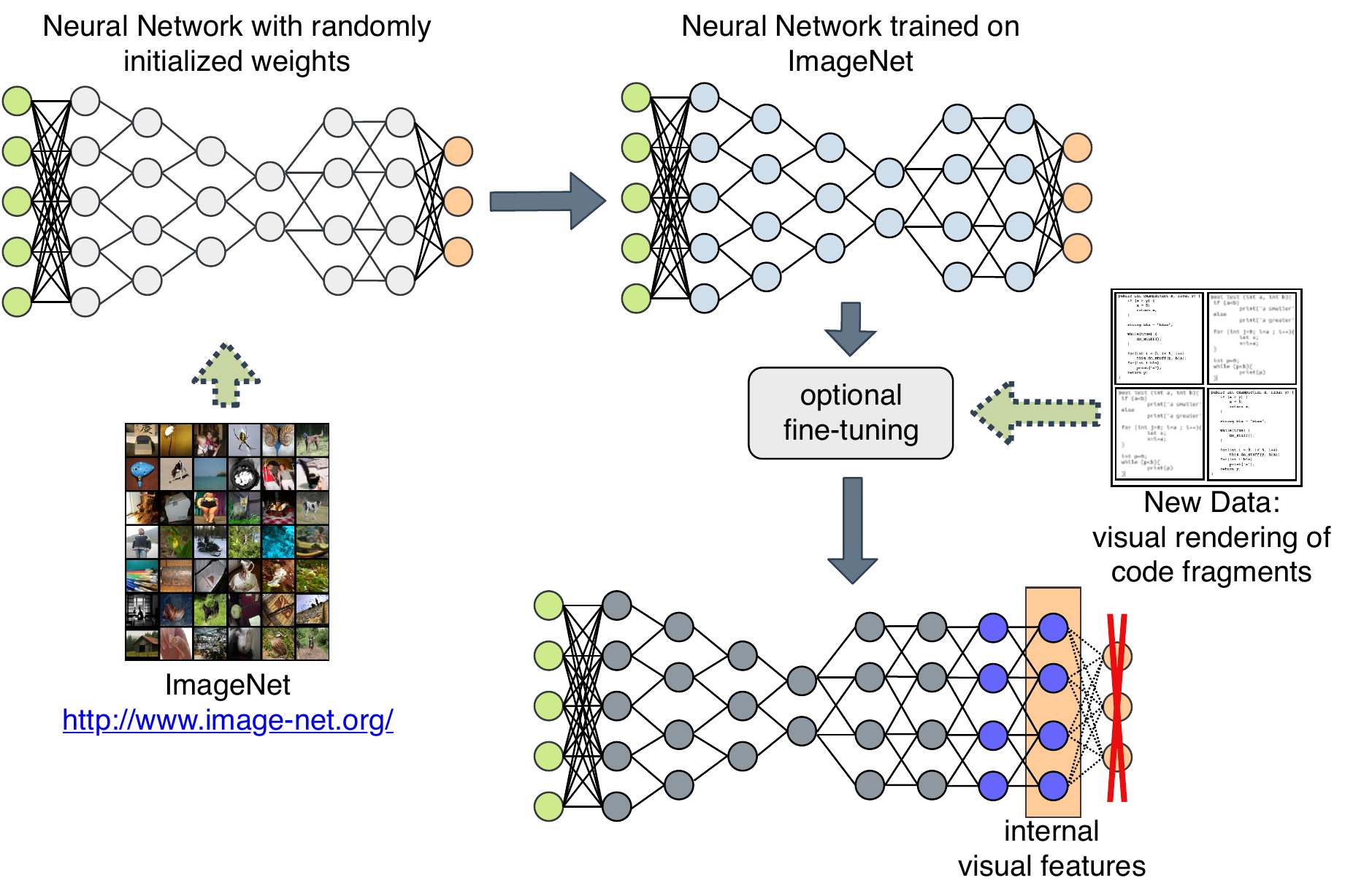}
    \caption{Principle of transfer learning applied to build our deep feature extractor for code}
    \label{fig:transfer-learning}
\end{figure}

\begin{figure*}[!t]
    \centering
    \includegraphics[width=\linewidth]{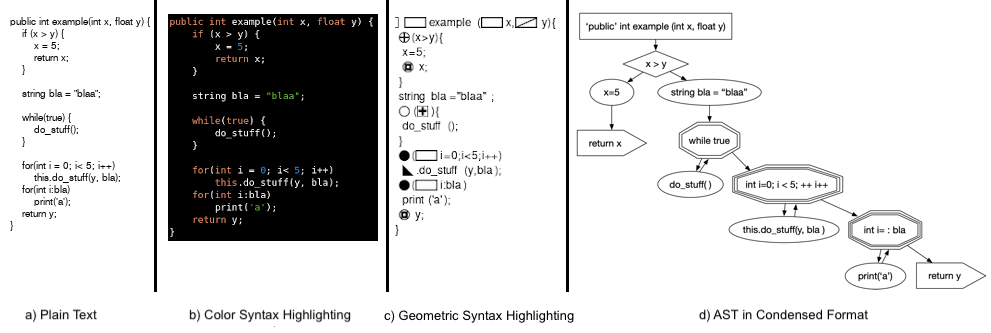}
    \caption{Variations of visual representations of code}
    \label{fig:visualisations}
\end{figure*}

Image classification neural networks consist of a multitude of convolutional layers that all learn different combinations and variations of the data contained in the previous layers. In addition, the networks have an input layer which accepts the input data and a fully connected output layer. 
This final layer is usually sized according to the number of possible labels and is in charge of deciding a label for the data coming from the previous layers. In our case, as depicted in Figure~\ref{fig:transfer-learning}, we focus on retrieving the intermediate features that are accessible in the penultimate layer. Actually, these features could have been collected on any previous layers. For the sake of prototyping speed, we immediately accessed the readily-available features. Future work could investigate other layers.

It should be noted that transfer learning is gaining traction within the deep learning community, since several domains lack sufficient data for training~\cite{pan2009survey}. Therefore, a fundamental motivation in the study of transfer learning is the fact that people can intelligently
apply knowledge learned previously to solve new problems
faster or with better solutions.
For example, it has been shown possible to use the knowledge about notes and rhythm, which were learned for playing the piano, to learn guitar playing; applying the vocabulary learned in French to infer English words as they share a certain base; or in audio-visual correspondence tasks~\cite{Arandjelovic_2017_ICCV}.



\subsection{Visualization options}
\label{subsec:visuals}
We explore in \toolname four variations of code visualizations in order to assess the influence of the selected visual representation on the performance of \toolname. We describe each visual representation by explaining its principle, detailing its implementation and arguing about its relevance. 

\noindent
$\bullet$ \textbf{{\sc Plain} Text:} The first visual representation is straightforward. It consists of simply rendering the textual representation of the code as a black and white image without highlighting any language construct. The rendering is implemented using the {\em pillow}\footnote{\url{https://pillow.readthedocs.io}} Python image drawing and manipulation library:  source code text is rendered as-is, i.e., with the indentations used by the developer, while applying a white background. {\sc Plain text}, illustrated in Figure~\ref{fig:visualisations}(a), is considered as our baseline visual representation of code.

\noindent
$\bullet$ \textbf{{\sc Color} Syntax Highlighting:} A simple variation of the {\sc Plain text} visualization consists in rendering the code text while highlighting syntax with colors, similarly to what is done in programming environments. This rendering approach is implemented by first generating an html page to highlight the code using the {\tt google code-prettify} javascript library. The web page is then saved as a PNG image using the {\em imgkit}\footnote{\url{https://pypi.org/project/imgkit/}}, a python wrapper for the Webkit web browser engine. As illustrated in Figure~\ref{fig:visualisations}(b), this visual representation is expliciting code structures for human programmers. Therefore, we expect that color-based syntax highlighting can be relevant for semantic machine learning tasks.

\noindent
$\bullet$ \textbf{{\sc Geometric} Syntax Highlighting:} 
In the previous visualization option, emphasis is put on color. Yet, image classification neural networks are also known to capture shapes. We propose to build a rendering of code where language keywords are represented by specific geometric shapes (i.e., icons). The implementation is based on the tokenization of code fragments using the {\em javalang}\footnote{\url{https://pypi.org/project/javalang/}} python library. We preset the mapping of language keywords with specific icons. During rendering, the text tokens are then replaced by the associated icons. Overall, although this representation could be nonsensical for humans, we expect that it will support the learning algorithm in the same way colored syntax does for programmers visual perception of code.

\noindent
$\bullet$ \textbf{{\sc Ast} in Condensed Format:}
Finally, we consider a visual rendering of the abstract syntax trees. The implementation is based on the AST generated by the {\em javalang} python library and leveraging  {\em graphviz}  python bindings\footnote{\url{https://www.graphviz.org/}}. 
To render the resulting graph, we generate a "graphviz" graph model by traversing the AST and representing some subtrees (e.g., the "for" loop control) in a purely textual manner, while representing other elements as their actual tree structure. This helps to condense the AST since raw AST quickly explodes in depth and breadth even for small code fragments.
In this representation we generated the graph such that the edges represent the possible control flows inside the code in order to capture its sequential nature. Further, we apply some geometric shapes to specific types of nodes in order to augment the visual strength of specific code structures. 

Overall, we try in \toolname visualization options that emphasize on colors, shapes and structures, and compare against the baseline plain text rendering. Although the generation of visual renderings is stable (i.e., not a random process), it should be noted that the AST in condensed format is, by far, the slowest to compute, as it involves many complex steps. 

Concretely, the output of the visualization rendering process is a single PNG image per visualization option and per code fragment. Each image may also be re-scaled to fit with the input requirements of the pre-trained image classification neural network. 


\subsection{Code classification architecture}
\label{subsec:architecture}

\begin{figure}[!t]
\centering
        \includegraphics[width=\linewidth]{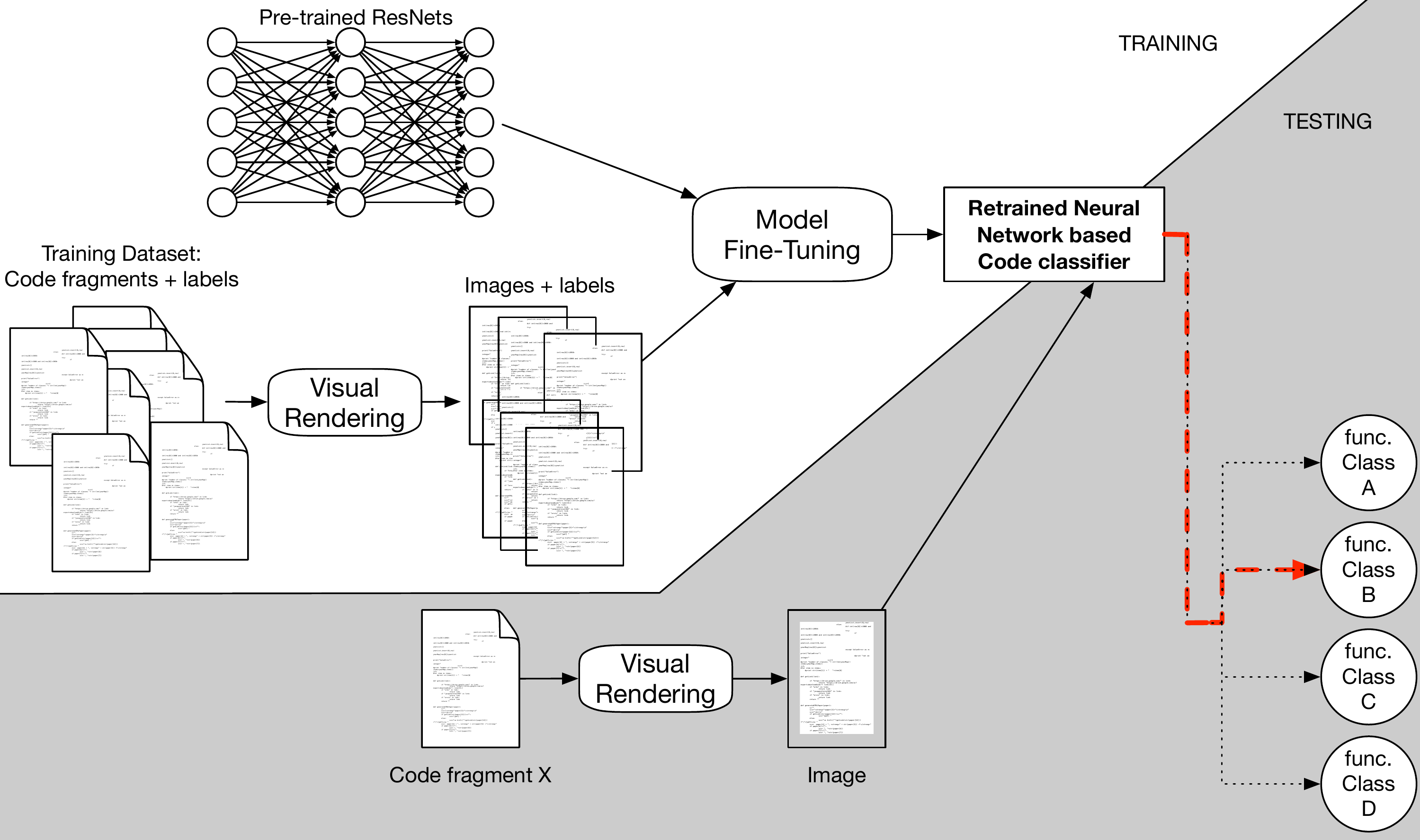}
\caption[Illustration of code classification]{Illustration of the architecture for \toolname's Code classification}
\label{fig:architecture-classification}
\end{figure}

Figure~\ref{fig:architecture-classification} provides a simple illustration of the overall architecture that we developed for code classification. We leverage neural networks (specifically, the powerful ResNets) that are pre-trained on the ImageNet dataset~\cite{deng2009imagenet}. However, we perform a re-training step, which is actually aimed to fine-tune the neural networks, towards better learning to extract features that are semantically-relevant to different classes of code functionalities. To that end, we update the size of the output layer of the pre-trained image classification network so that the final size accounts also for the number of possible functionality labels in our code dataset. The re-trained (i.e., fine-tuned) network on the training dataset is then used as classifier to predict the labels of code fragments in the test set based on their visual renderings. Since we do not tune any hyper-parameters of the network, we do not need a validation set. Nevertheless, we ensure that the process of re-training is performed for an empirically determined number of epochs\footnote{We ran a few experiments to check when the results stabilize.}. Details on how training and test sets are split are provided and discussed later in Section~\ref{sec:datasets}.

\subsection{Clone detection architecture}
\begin{figure}[!t]
\centering
        \includegraphics[width=\linewidth]{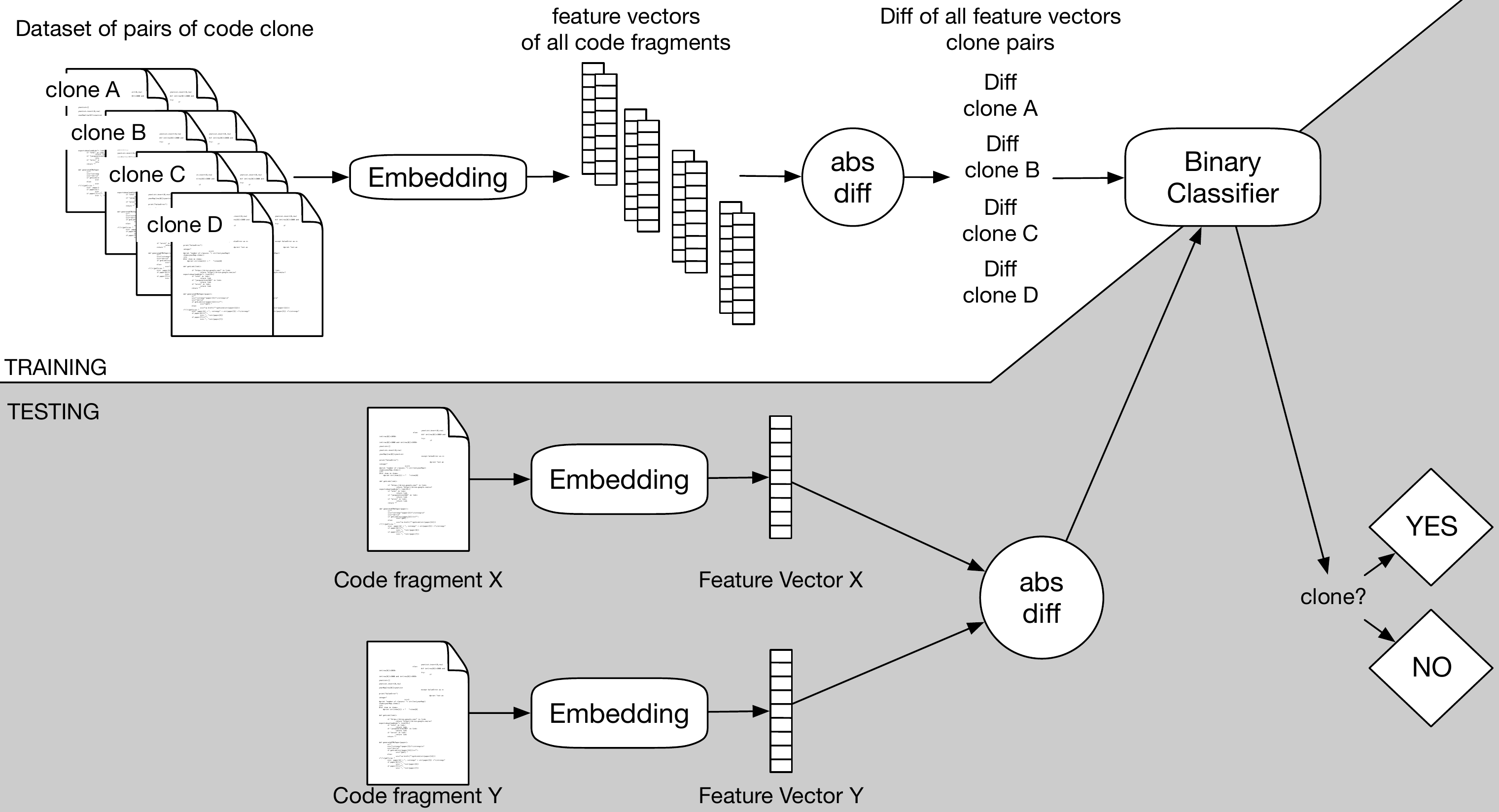}
\caption[Illustration of clone detection]{Illustration of the architecture for \toolname's Clone detection}
\label{fig:architecture-detection}
\end{figure}

Figure~\ref{fig:architecture-detection} illustrates \toolname's architecture for clone detection. Similarly to the pipeline of code classification, we leverage a pre-trained neural network for visual classification to which we feed the images obtained from visual renderings of code snippets. In this case, however, our objective is to simply collect the embeddings produced during deep feature extraction. Thus, given that we do not need the network to learn about new classes in our new (code-related) image datasets, we propose to directly use the ResNets that were pre-trained on ImageNet datasets. We expect the embeddings to still be relevant for capturing structural features. The feature vectors (i.e., embeddings) are then used to train binary classifiers as in traditional machine learning. Concretely, to train the binary classifiers, the first step is to calculate the absolute difference between the feature vectors of the candidate pair vectors. Those difference vectors can then be used to train our binary classifiers.


\subsubsection*{Algorithms for binary classification}
In code classification, we directly reuse the in-built capability of the neural networks to perform classification (i.e., using the {\em softmax} activation function at the last layer). Indeed, given that the input of the task is a single image representing the visualization rendering of a code fragment, the classical image classification neural network is suitable.

In clone detection, however, the input is a pair of code fragments (precisely, a pair of images taken from their visualization renderings). This means that the architecture of image classification networks is not readily applicable for this case as it always expects a single input only. For sake of simplicity and optimization, we decided to use the neural network to collect embeddings for individual images, and train our final classifier separately. This strategy allows us to experiment with different traditional classification algorithms. Our experiments provide results with Support Vector Machines \cite{cortes1995support}, k-Nearest Neighbours~\cite{cover1967nearest} and a simple binary classification neural network~\cite{haykin1994neural}.
\newpage
\section{Experimental Setup}
We enumerate the research questions, overview the datasets used in the experiments and discuss some important implementation details. We open-source the implementation of our prototype implementation of \toolname and release all data related to the experiments recorded in this paper. The artifact web page is currently in an anonymous repository: \url{https://github.com/wysiwim/wysiwim}

\subsection{Research Questions}
\begin{description}
\item \textbf{RQ1: How does \toolname perform in comparison with the state-of-the-art?} We investigate the ability of our novel approach of semantics learning based on visual representation of code to keep up with the state-of-the-art for the tasks of code classification and clone detection.

\item \textbf{RQ2: How does the visual representation influence the performance of \toolname?} Experiments for this research question are focused on the code clone detection task, where we try all the considered visual representations options and compare the performance differences.

\item \textbf{RQ3: What is the impact of the classification algorithms on \toolname?} We investigate in this research question different supervised learning algorithms that can be leveraged to train the binary classifiers needed for the code clone detection architecture.
\end{description}

\subsection{Selection of datasets}
\label{sec:datasets}
\paragraph{\textbf{Code Classification:}}
We assess the performance of \toolname for the code classification task based on the Open Judge (OJ) dataset as introduced in~\cite{mou2016convolutional}. This choice is motivated by the need to directly compare against the state-of-the-art (namely, TBCNN~\cite{mou2016convolutional} and ASTNN~\cite{zhang2019novel}), which also run experiments on this dataset.
This dataset contains 104 different functionalities and 500 samples per functionality.
In order to achieve balanced datasets for training and testing, we apply a stratified sampling over the functionalities with a ratio of 4:1 (i.e, 80\% of data for training and 20\% for testing).

\paragraph{\textbf{Code Clone Detection:}}
The state-of-the-art for code clone detection being ASTNN~\cite{zhang2019novel}, we reuse the dataset that they release in their experiment artifacts. This enables a direct and unbiased comparison. The split into training and testing sets is also predefined and applied as-is.
This dataset consists of 20k Type-4 clone pairs and 20k non-clone pairs.

Nevertheless, we found that the ASTNN dataset is not balanced with respect to the number of clones per functionality. Thus we selected a custom subset of BigCloneBench (BCB) (cf. Section~\ref{subsubsec:benchmarks}) for our further experiments. We focus on code fragments related to three functionalities (i.e.,  \#7 - {\tt bubble-sort array} \#13 - {\tt  shuffle array inplace} and \# 44 - {\tt check for palindrome}) which we  consider the most suitable for our evaluation: these code fragments are dissimilar enough but concise; furthermore \#7 and \#13 code fragments deal all with arrays and yet semantically distant, offering an opportunity to properly assess the semantic clone detection approach.
The dataset is constructed by randomly sampling 500 Type-4 clone pairs and 500 non-clone pairs per functionality. The ground truth information of clone/non-clone is based on the annotations  provided in BigCloneBench.

\subsection{Implementation}
Our proof-of-concept implementation of \toolname is written in Python using common frameworks and libraries. In particular, several Python libraries are leveraged for the code fragment processing towards producing visual renderings as images (cf. Section~\ref{subsec:visuals}). We leverage the {\em pandas}\footnote{\url{https://pandas.pydata.org/}} library for data management. 
Further for the stratified splitting of the datasets, we use the dataset splitting method from the {\em scikit-learn}\footnote{\url{https://scikit-learn.org/}} library.
\paragraph{{Data Preprocessing}}
The pre-trained networks considered in our experiments have a limitation on the input image size being set to exactly 224x224 pixels. To fit with this requirement, we choose to simply re-scale the code visualizations to this size.

\paragraph{Data Augmentation.} Usually in many machine learning applications, a data augmentation step is performed. In image classification in particular, one usually uses a set of random transformations to create many variations from the input data in order to artificially increase the size of the dataset. Those random transformations include rescaling, cropping, mirroring etc.
For our approach however, we found that this is not beneficial since source code naturally does not appear up-side down or mirrored.

\paragraph{\textbf{Code classification}}
The implementation of our code classification task is mainly based on PyTorch~\cite{paszke2017pytorch} and uses the pre-trained ResNet models provided by the PyTorch framework. In particular we use a ResNet18 and a ResNet50 to highlight the increase of performance when the number of layers is increased. Both are pre-trained on the ImageNet dataset~\cite{deng2009imagenet}.

\paragraph{\textbf{Clone detection}}
For the implementation of the binary clone detection task, we use again the pre-trained ResNet50 model and drop the last layer in order to generate the raw feature vectors for our visualized code fragments. Those vectors are then converted into {\em numpy}\footnote{\url{https://www.numpy.org/}} arrays which facilitates the calculation of the absolute difference between vectors. For the final stage of learning and predicting, we use pytorch again to implement a simple binary classification network. SVM and k-NN algorithm implementations are taken from the {\em scikit-learn} library.


\section{Results}
We now present the experimental results in response to the research questions, and based on the experimental settings presented previously.

\subsection{RQ1: [ Performance of \toolname]}
\subsubsection{Code classification}
For performance comparison against the state-of-the-art for the task of code classification, we focus on the {\bf accuracy} metric, which is used by the state-of-the-art ASTNN and TBCNN authors to report their performance (see.~\citep{zhang2019novel,mou2016convolutional}). As discussed previously, we also reuse the same OJ dataset that was used for ASTNN and TBCNN validation. We apply the {\sc Plain} text visualization to render code.

\textbf{Results:} Table~\ref{tab:results-rq1-cc} provides the accuracy metrics of different approaches. \toolname provides an accuracy of 89.7 and 86.4 percent for code classification on the OJ dataset with ResNet18 and ResNet50 respectively. These results suggest that  we perform reasonably well in comparison to the state-of-the-art which are reported to yield accuracy scores of 94.0\% and 98.2\% for TBCNN and ASTNN respectively. Given the limitations that our implementation carries (cf. Section~\ref{subsec:limits}) and the potential for improvement (cf. Section~\ref{subsec:lessons}), we view this performance of \toolname as largely promising.

\begin{table}[!h]
\begin{tabular}{@{}lcc@{}}
\toprule
Method             & Variation        & Accuracy \\ \midrule
TBCNN              &    -     & 94.0     \\ 
ASTNN             &     -     & 98.2     \\ \midrule
\toolname  & with ResNet18 pre-trained model & 86.4     \\
\toolname  & with ResNet50 pre-trained model & 89.7     \\ \bottomrule
\end{tabular}
\caption[Results RQ1 - Code Classification: Comparison to SOTA]{Accuracy comparisons for code classification.}
\label{tab:results-rq1-cc}
\vspace{-0.5cm}
\end{table}
\vspace{-0.5cm}

\subsubsection{Clone Detection}
Experiments for Clone detection are done with the BigCloneBench which already have labels on pairs of clones and non-clones. For fair comparisons, we run ASTNN and \toolname on the same samples of Type-4 clones that were used to evaluate ASTNN by Zhang et al.~\cite{zhang2019novel}. For this experiment, we present the results for our best configurations: the {\sc Ast} in condensed format as the visualization option and the neural network binary classification algorithm. We refer the reader to next experiments where we show that the visualization and algorithms have a limited impact on the performance of \toolname. Finally, contrary to previous experiments, we do not compare against TBCNN since this approach has not been applied for clone detection.

\textbf{Results:} The results provided in Table~\ref{fig:results-rq1-ccd} show that, overall, we perform similarly well as the state-of-the-art in terms of F-Measure. It is further noteworthy that \toolname offers a better trade-off between precision and recall than ASTNN which present quasi-perfect precision but lower recall. 

\begin{table}[!h]
\begin{tabular}{@{}llccc@{}}
\toprule
Method        & F1 score & Precision & Recall \\ \midrule
ASTNN    & 93.7     & 99.8      & 88.3   \\
\toolname    & 94.8     & 95.4      & 94.3   \\ \bottomrule
\end{tabular}
\caption[Results RQ1: Clone Detection Comparison to SOTA]{Performance comparison for clone detection.}
\label{fig:results-rq1-ccd}
\vspace{-0.5cm}
\end{table}

\vspace{-0.5cm}

\subsection{RQ2: [ Visualization influence ]}
To examine the influence of visualization rendering options, we consider the clone detection task where \toolname implements the binary neural network classifier for the final clone decision. The process is then performed for all previously-described visual representations options (cf. Section~\ref{subsec:visuals}).

\begin{figure}[!h]
\centering
        \includegraphics[width=0.7\linewidth]{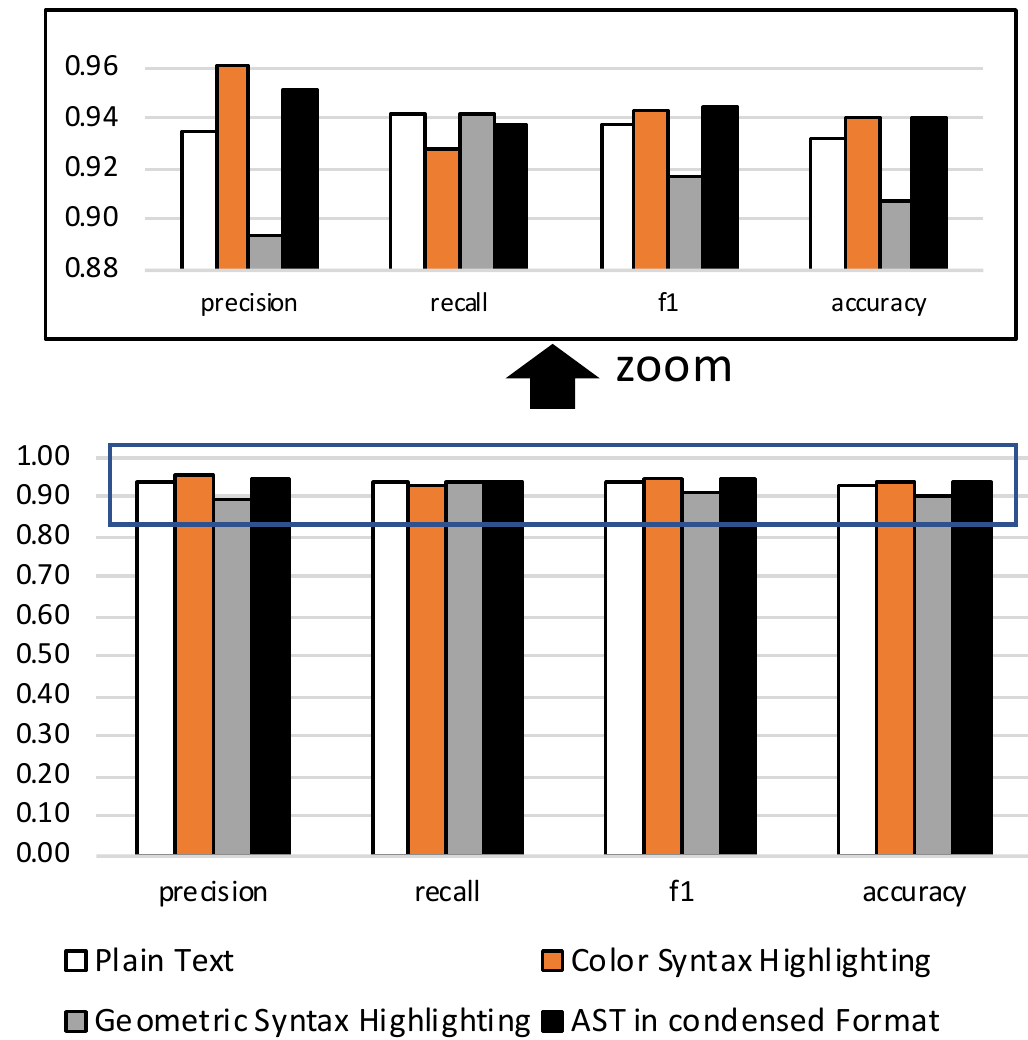} 
\caption[Results RQ2: Influence of Visualization]{Influence of visual rendering schemes on clone detection performance.}
\label{fig:results-rq2}
\vspace{-0.5cm}
\end{figure}

\textbf{Results:} The results depicted in figure \ref{fig:results-rq2} suggest that the {\sc Ast} in condensed format and the {\sc Color} syntax highlighting visual representations yield the best results (which are further similar for these two representations).

On the one hand, it is noteworthy that the {\sc Color} syntax highlighting improves over the {\sc Plain} text visualization, hence confirming our initial intuition that colors can help to better capture semantics visually. On the other hand, although {\sc Geometric} syntax highlighting performs slightly less well than others, it's relatively high performance indeed suggests that visual shapes are expressive enough to help learn semantics of code structures. In any case, we also suspect that the performance degradation of {\sc Geometric} syntax highlighting visualizations might emerge from a bad choice of the keyword substitution shapes. Finally, we note that, depending on the performance metric, any of the visualizations may perform better or worse than other visualizations.


\subsection{RQ3: [ Algorithm impact ]}
To run several experiments of clone detection while varying the classification algorithms, we leverage our main dataset sampled from BCB (cf. Section~\ref{sec:datasets}). We also fix the visualization option to the {\sc Ast} condensed format. The experiments are then performed to compare the variations of sensitivity of the \toolname embeddings with respect to different algorithms. We use kNN\footnote{We use kNN with the default setting of scikit-learn where $k=5$}, a simple NN\footnote{The NN is a simple fully-connected linear layer with bias, so in essence a linear combination}, and SVM. 

\begin{figure}[!h]
\centering
        \includegraphics[width=0.7\linewidth]{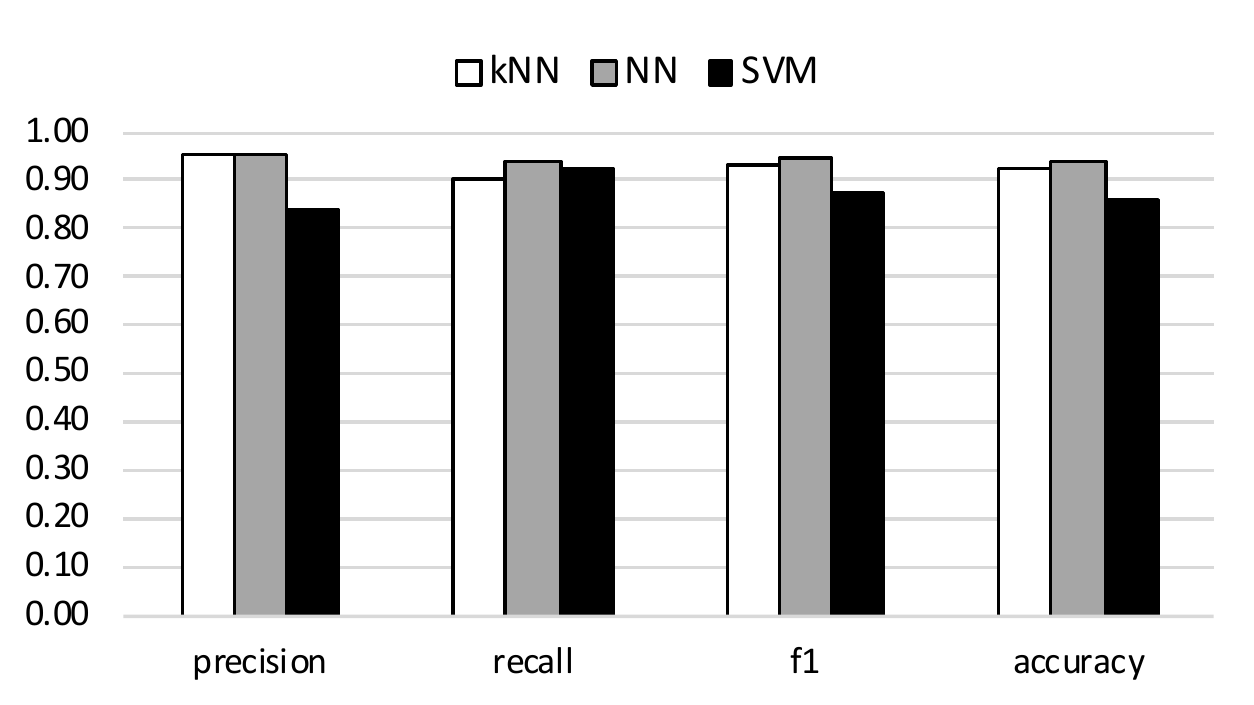} 
\vspace{-0.3cm}
\caption[Results RQ3: Influence of classification algorithm]{Impact of classification algorithms on clone detection performance}
\label{fig:results-rq3}
\vspace{-0.5cm}
\end{figure}

\textbf{Results:} Figure~\ref{fig:results-rq3} presents the comparison results. It appears that the algorithm has a slight impact on recall scores between kNN and the Neural Network classifier, while they yield the same precision. In contrast the precision of SVM\footnote{we use the default Support Vector classifiers implementation in scikit-learn without tuning any parameters} is lower by 8 percentage points. Nevertheless, all three algorithms offer reasonably good performance, which suggests that the embeddings produced by the pre-trained models are effective in terms of semantic representations.

\section{Discussion}
\label{sec:discussion}
Our experimental evaluation bears some threats to validity, while the approach itself has limitations that can be improved in future work.
\vspace{-0.5cm}
\subsection{Threats to Validity}
\noindent
{\textbf{Internal Validity - Dataset.}} Our dataset selections are limited in terms of size and diversity of functionalities. For the clone detection variant in particular, even though the number of clones is rather high, the number of code fragments that the clone and non-clones pairs are based on is still very small as the pairs are formed from those base code fragments by pairwise combination. This lack of diversity might negatively influence the generalizability of the evaluation results. Nevertheless, we mitigate this threat in the comparison experiment (RQ1) by using the same datasets as the state-of-the-art (i.e., ASTNN).

\noindent{\textbf{External Validity - Dataset.}} Even though the BigCloneBench is widely used throughout the literature, the judgment of whether or not a pair of code fragments form a clone remains biased and purely based on benchmark authors' intuition. Further, there is no single or precise notion of what semantic similarity is. Thus, the semantic boundaries of the functionality classes might not be consistent across all the represented functionalities. Finally, the program semantics of a code fragment might be obscured by the usage of external libraries that are not included within the dataset, in which case the decision task is technically unfeasible.

\noindent{\textbf{External Validity - Presence of clone duplicates in BCB.}} During development, we noticed the existence of conceptually duplicated clones in  BigCloneBench. This fact showed up in the form of identical visual representations of code for different code fragment ids. It turned out that those fragments emerged from Type-1 clone pairs, which are technically the same code.
When those both clone fragments are combined with another code fragment to form clone pairs, those clone pairs are conceptually duplicated. 
Although we cannot provide precise statistics on the extent of clone duplicates present in  BigCloneBench, we can approximate, based on the code fragments that are used in Type-1 and Type-4 clone pairs, an upper-bound of approximately 30 percent clone duplicates. If we consider that Type-2 and Type-3 code snippets can also build clone duplicates, this estimation goes up to even 60 percent.
In our case, during the development, we experienced drops of performance of about 10 percent, on small development examples. Hence we conclude that code clones should not be disregarded if precise and valid evaluations are desired. This conclusion is consistent with recent empirical results reported by Alamanis on the adverse effects of code duplication in machine learning models of code~\cite{allamanis2018adverse}.

\begin{figure}[!h]
\centering
        \includegraphics[width=0.9\linewidth]{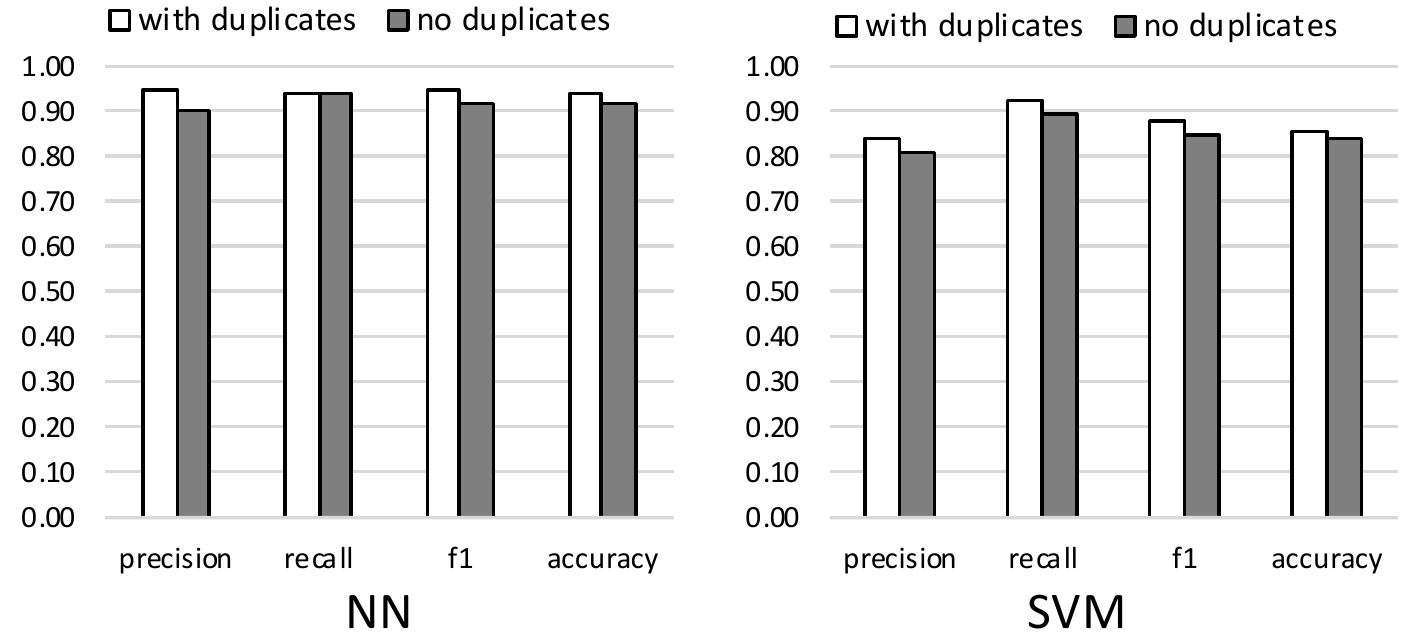}
\caption[Results RQ4: Influence of clone duplicates]{Influence of clone duplicates.}
\label{fig:results-rq4}
\vspace{-0.3cm}
\end{figure}

To explore the impact of code duplicates on the performance of \toolname, we build a dataset (based on the same three functionalities and numbers of clones/non-clone pairs) where we do not 
use any clones that contain code fragments that are also used in Type-1 clone pairs. The results from figure \ref{fig:results-rq4} show that the avoidance of clone duplicates slightly degrades the overall results. This makes sense since the existence of clone duplicates makes the task easier and allows to achieve a higher score. This finding is further valid for both the best-performing algorithm (NN) and the worst-performing one (SVM).

\noindent{\textbf{Construct Validity - Dataset.}}
A recurrent construct validity issue in the machine learning literature is related to class imbalance. In clone detection, one must ensure that all functionalities are balanced in the dataset of clone and non-clone pairs. Some approaches may overfit to specific (and largely represented) classes. To check for this issue, we build a balanced dataset (with and without duplicates) and compared the performance of \toolname clone detection on this dataset as well as the imbalanced dataset provided in ASTNN artifacts. Indeed, the ASTNN dataset is randomly sampled from the BCB (using a fixed random seed) and hence -more or less- keeps the unbalancing that is present in the BCB itself. Comparison
results in Figure~\ref{fig:results-balanced} with balanced and imbalanced (i.e., ASTNN dataset) suggest that \toolname keeps its promises on performance. 

\begin{table}[!h]
\begin{tabular}{@{}llccc@{}}
\toprule
Dataset                    &  Accuracy &    F1 score &  Precision     & Recall  \\ \midrule
ASTNN                      & 91.8      & 94.8        &     95.4  & 94.3 \\\midrule
 balanced                  & 94.1      & 94.5        &     95.2   & 93.8 \\
balanced w/o duplicates    &  92.1     &  92.0       &     94.1  & 90.1 \\
  \bottomrule

\end{tabular}
\caption[Further results: balanced dataset]{Impact of class imbalance in the dataset of code clones}
\label{fig:results-balanced}
\vspace{-0.5cm}
\end{table}

\noindent{\textbf{Construct Validity - Cross validation.}} We did not perform any cross validation on our approach, as our goal was to rather convey the concepts behind the approach rather than achieve high results. It is probable that the exact results vary to a certain extent on different splits of the dataset, especially since the different code fragments are probably not "semantically equally diverse" to each other, without further specifying what that could mean.

\noindent{\textbf{Conclusion Validity - Lack of definitions of semantic similarity.}} The software engineering community faces a crucial challenge for defining what semantic similarity means. Since we do not dare to explicitly define what semantic similarity means, we have to rely on the semantic value that is embedded in our dataset, respectively as it was implied by the creators of the BCB. In consequence, a specific selection of a subset of the dataset may even influence the overall semantics it carries. However, even when two approaches are applied on the same dataset, they might still view semantic similarity differently. These facts make it hard to evaluate and especially compare semantic approaches of any sort.

\subsection{Limitations}
\label{subsec:limits}
\noindent
{\textbf{Input size of ResNets.}} Image classification networks have technically and by construction a strong limitation on their input size. This is problematic as it introduces loss and distortion of our input data. In consequence, we may completely loose the fine-grained lexical information that is contained in the visual representations of our code fragments. 

\noindent{\textbf{Code fragment granularity.}} The approach as presented is mainly designed to work with method granularity code fragments. Image classification networks are designed to assign a single most suitable label to a whole single input. This is consistent with generally accepted good coding style rules, which claim that a single method should always implement a single functionality (known as the single responsibility principle) ~\cite{martin2000design}. To enlarge the scope of the granularity, our core concept of code visualization could be leveraged to full programs by applying object localization instead to detect what functionalities a software is composed of. This principle could also explain why our approach works slightly less well on the OJ dataset, which consists of whole programs, while the BigCloneBench rather provides method level granularities.

\noindent{\textbf{Colors in visualizations.}} Our visualizations apply colors only very sparsely, in the case of the color syntax highlighting variant, or not at all for the other visual representations. The current implementation of \toolname is thus not fully leveraging the potential of ResNet, which is designed to operate on all 3 color channels.  

\noindent{\textbf{Traditional classification algorithms.}} For the clone detection task  we apply very basic binary classification algorithms. These algorithms do probably not explore all semantics learned by the ResNet deep feature extractor.

\noindent{\textbf{Scope of the clone datasets.}} The datasets are not only a threat to validity but also a major limitation. Our hypothesis is that, due to the limited variety and size of the datasets available today, it is not possible yet to learn general semantic knowledge that can by applied to all possible data.

\subsection{Lessons learned and Future work}
\label{subsec:lessons}
As the current implementation of \toolname represents only a proof-of-concept with limited goals, it offers a lot of potential for extensions and improvements. Furthermore, the general concept of visualizing code and learning on those visual representations could be interesting also to other software engineering tasks, or could be combined with existing approaches. Beyond our approach, we identified some general current limitations on the task of semantic code clone detection, such as the lack of suitable datasets and benchmarks but also the lack of more precise and actionable definitions of semantics or semantic similarity.
\paragraph{{Mitigating Image classifier input limitation}} As mentioned in the previous sub-section, a major limitation of our approach is the fixed input image size of the ResNet classifier. One potential way to mitigate this limitation could be to slice the image into multiple images of the required input size. Those slices could then be used to generate a larger feature vector, representing the whole image. This  would allow to capture more fine-grained information as well. Of course, it might be necessary to apply also scaled versions of the images to capture large-scale structural information too.
\paragraph{{Visualizations}}
As our visualizations showed, the use of colors can have a positive effect on the results. However, as our condensed AST visualization yielded the best overall results, it might be interesting to further apply color coding on ASTs to make better use of the full potential of the image classification neural networks.
\paragraph{{Datasets and Benchmarks:}} A future work that is important beyond our approach is the development of datasets and benchmarks that are more suitable for semantic code clone detection and semantic approaches in general. This includes a high number of different functionalities and a high number of diverse code examples per functionality. Especially sets providing a multitude of more basic functionalities that do not depend on external libraries would be desirable. They would allow to learn models the way humans learn semantics of computing languages, by starting very small.

\paragraph{{Data augmentation:}} Similarly to data augmentation done in image classification via generating variant images through rotation, cropping, etc., we could envision to apply a data augmentation, although at the meta level, such as mutating the code in semantically-equivalent ways in order to increase the size of our dataset.

\paragraph{{Actionable definitions of semantics (similarity):}} Another very important future work would be to make efforts towards actionable definitions of semantics, or semantic similarity. A possible approach to this could be the definition of semantics through software tests. As software tests represent an executable variant of software specifications, they give a good notion of the requirements we put into our semantics. Of course, there are a few problematic aspects in this approach. One aspect is that each application may require different abstractions of a certain functionality. Another aspect is that the code snippets for a certain functionality would all have to use test suites.

\section{Conclusions}
\label{sec:conclusion}
We presented a novel direction to code semantics learning based on visualization and transfer learning. \toolname exploits the power of pre-trained ResNets to extract deep features from visualization renderings of source code fragments. We apply this approach to two variants of clone identification, namely code classification and clone detection. Experimental results on BigCloneBench and the Open Judge datasets show that our approach performs reasonably well and can keep up with the state-of-the-art within the scope of our experimental settings (which we carefully design to be comparable to literature experiments). 
Our experiments reveal that visualizations  of AST yield the best overall clone detection results. We complete the paper by enumerating a list of limitations, which, if resolved, may unleash a huge potential of \toolname beyond clone identification tasks. 

\noindent
{\bf Availability: } All experimental data as well as the source code of \toolname is open sourced in an anonymous repository:
\begin{center}
    \url{https://github.com/wysiwim/wysiwim}
\end{center}



\balance
\bibliographystyle{ACM-Reference-Format}
\bibliography{article.bib}


\begin{thebibliography}{48}


\ifx \showCODEN    \undefined \def \showCODEN     #1{\unskip}     \fi
\ifx \showDOI      \undefined \def \showDOI       #1{#1}\fi
\ifx \showISBNx    \undefined \def \showISBNx     #1{\unskip}     \fi
\ifx \showISBNxiii \undefined \def \showISBNxiii  #1{\unskip}     \fi
\ifx \showISSN     \undefined \def \showISSN      #1{\unskip}     \fi
\ifx \showLCCN     \undefined \def \showLCCN      #1{\unskip}     \fi
\ifx \shownote     \undefined \def \shownote      #1{#1}          \fi
\ifx \showarticletitle \undefined \def \showarticletitle #1{#1}   \fi
\ifx \showURL      \undefined \def \showURL       {\relax}        \fi
\providecommand\bibfield[2]{#2}
\providecommand\bibinfo[2]{#2}
\providecommand\natexlab[1]{#1}
\providecommand\showeprint[2][]{arXiv:#2}

\bibitem[\protect\citeauthoryear{Allamanis}{Allamanis}{2018}]%
        {allamanis2018adverse}
\bibfield{author}{\bibinfo{person}{Miltiadis Allamanis}.}
  \bibinfo{year}{2018}\natexlab{}.
\newblock \showarticletitle{The Adverse Effects of Code Duplication in Machine
  Learning Models of Code}.
\newblock \bibinfo{journal}{\emph{arXiv preprint arXiv:1812.06469}}
  (\bibinfo{year}{2018}).
\newblock


\bibitem[\protect\citeauthoryear{Alon, Brody, Levy, and Yahav}{Alon
  et~al\mbox{.}}{2018}]%
        {alon2018code2seq}
\bibfield{author}{\bibinfo{person}{Uri Alon}, \bibinfo{person}{Shaked Brody},
  \bibinfo{person}{Omer Levy}, {and} \bibinfo{person}{Eran Yahav}.}
  \bibinfo{year}{2018}\natexlab{}.
\newblock \showarticletitle{code2seq: Generating sequences from structured
  representations of code}.
\newblock \bibinfo{journal}{\emph{arXiv preprint arXiv:1808.01400}}
  (\bibinfo{year}{2018}).
\newblock


\bibitem[\protect\citeauthoryear{Alon, Zilberstein, Levy, and Yahav}{Alon
  et~al\mbox{.}}{2019}]%
        {alon2019code2vec}
\bibfield{author}{\bibinfo{person}{Uri Alon}, \bibinfo{person}{Meital
  Zilberstein}, \bibinfo{person}{Omer Levy}, {and} \bibinfo{person}{Eran
  Yahav}.} \bibinfo{year}{2019}\natexlab{}.
\newblock \showarticletitle{code2vec: Learning distributed representations of
  code}.
\newblock \bibinfo{journal}{\emph{Proceedings of the ACM on Programming
  Languages}} \bibinfo{volume}{3}, \bibinfo{number}{POPL}
  (\bibinfo{year}{2019}), \bibinfo{pages}{40}.
\newblock


\bibitem[\protect\citeauthoryear{{Ambient Software Evoluton Group}}{{Ambient
  Software Evoluton Group}}{2013}]%
        {ijadata}
\bibfield{author}{\bibinfo{person}{{Ambient Software Evoluton Group}}.}
  \bibinfo{year}{2013}\natexlab{}.
\newblock \bibinfo{title}{{IJaDataset} 2.0, http://secold.org/ \\
  projects/seclone}.
\newblock
\newblock


\bibitem[\protect\citeauthoryear{Arandjelovic and Zisserman}{Arandjelovic and
  Zisserman}{2017}]%
        {Arandjelovic_2017_ICCV}
\bibfield{author}{\bibinfo{person}{Relja Arandjelovic} {and}
  \bibinfo{person}{Andrew Zisserman}.} \bibinfo{year}{2017}\natexlab{}.
\newblock \showarticletitle{Look, Listen and Learn}. In
  \bibinfo{booktitle}{\emph{The IEEE International Conference on Computer
  Vision (ICCV)}}.
\newblock


\bibitem[\protect\citeauthoryear{Baker}{Baker}{1992}]%
        {baker1993program}
\bibfield{author}{\bibinfo{person}{B.S. Baker}.}
  \bibinfo{year}{1992}\natexlab{}.
\newblock \showarticletitle{A Program for Identifying Duplicated Code}. In
  \bibinfo{booktitle}{\emph{Computing Science and Statistics: Proceedings of
  the 24th Symposium on the Interface}}, Vol.~\bibinfo{volume}{24}.
  \bibinfo{pages}{49--57}.
\newblock
Issue Mar.


\bibitem[\protect\citeauthoryear{Baxter, Yahin, Moura, Sant'Anna, and
  Bier}{Baxter et~al\mbox{.}}{1998}]%
        {baxter1998clone}
\bibfield{author}{\bibinfo{person}{Ira~D Baxter}, \bibinfo{person}{Andrew
  Yahin}, \bibinfo{person}{Leonardo Moura}, \bibinfo{person}{Marcelo
  Sant'Anna}, {and} \bibinfo{person}{Lorraine Bier}.}
  \bibinfo{year}{1998}\natexlab{}.
\newblock \showarticletitle{Clone detection using abstract syntax trees}. In
  \bibinfo{booktitle}{\emph{Proceedings of the International Conference
  onSoftware Maintenance}}. IEEE, \bibinfo{pages}{368--377}.
\newblock


\bibitem[\protect\citeauthoryear{Bellon, Koschke, Antoniol, Krinke, and
  Merlo}{Bellon et~al\mbox{.}}{2007}]%
        {bellon2007comparison}
\bibfield{author}{\bibinfo{person}{Stefan Bellon}, \bibinfo{person}{Rainer
  Koschke}, \bibinfo{person}{Giulio Antoniol}, \bibinfo{person}{Jens Krinke},
  {and} \bibinfo{person}{Ettore Merlo}.} \bibinfo{year}{2007}\natexlab{}.
\newblock \showarticletitle{Comparison and evaluation of clone detection
  tools}.
\newblock \bibinfo{journal}{\emph{IEEE Transactions on Software Engineering}}
  \bibinfo{volume}{33}, \bibinfo{number}{9} (\bibinfo{year}{2007}),
  \bibinfo{pages}{577--591}.
\newblock


\bibitem[\protect\citeauthoryear{Ben-Nun, Jakobovits, and Hoefler}{Ben-Nun
  et~al\mbox{.}}{2018}]%
        {ben2018neural}
\bibfield{author}{\bibinfo{person}{Tal Ben-Nun},
  \bibinfo{person}{Alice~Shoshana Jakobovits}, {and} \bibinfo{person}{Torsten
  Hoefler}.} \bibinfo{year}{2018}\natexlab{}.
\newblock \showarticletitle{Neural code comprehension: a learnable
  representation of code semantics}. In \bibinfo{booktitle}{\emph{Advances in
  Neural Information Processing Systems}}. \bibinfo{pages}{3585--3597}.
\newblock


\bibitem[\protect\citeauthoryear{Chen and Monperrus}{Chen and
  Monperrus}{2019}]%
        {chen2019literature}
\bibfield{author}{\bibinfo{person}{Zimin Chen} {and} \bibinfo{person}{Martin
  Monperrus}.} \bibinfo{year}{2019}\natexlab{}.
\newblock \showarticletitle{A Literature Study of Embeddings on Source Code}.
\newblock \bibinfo{journal}{\emph{arXiv preprint arXiv:1904.03061}}
  (\bibinfo{year}{2019}).
\newblock


\bibitem[\protect\citeauthoryear{Cortes and Vapnik}{Cortes and Vapnik}{1995}]%
        {cortes1995support}
\bibfield{author}{\bibinfo{person}{Corinna Cortes} {and}
  \bibinfo{person}{Vladimir Vapnik}.} \bibinfo{year}{1995}\natexlab{}.
\newblock \showarticletitle{Support-vector networks}.
\newblock \bibinfo{journal}{\emph{Machine learning}} \bibinfo{volume}{20},
  \bibinfo{number}{3} (\bibinfo{year}{1995}), \bibinfo{pages}{273--297}.
\newblock


\bibitem[\protect\citeauthoryear{Cover, Hart, et~al\mbox{.}}{Cover
  et~al\mbox{.}}{1967}]%
        {cover1967nearest}
\bibfield{author}{\bibinfo{person}{Thomas~M Cover}, \bibinfo{person}{Peter
  Hart}, {et~al\mbox{.}}} \bibinfo{year}{1967}\natexlab{}.
\newblock \showarticletitle{Nearest neighbor pattern classification}.
\newblock \bibinfo{journal}{\emph{IEEE transactions on information theory}}
  \bibinfo{volume}{13}, \bibinfo{number}{1} (\bibinfo{year}{1967}),
  \bibinfo{pages}{21--27}.
\newblock


\bibitem[\protect\citeauthoryear{Deng, Dong, Socher, Li, Li, and Fei-Fei}{Deng
  et~al\mbox{.}}{2009}]%
        {deng2009imagenet}
\bibfield{author}{\bibinfo{person}{Jia Deng}, \bibinfo{person}{Wei Dong},
  \bibinfo{person}{Richard Socher}, \bibinfo{person}{Li-Jia Li},
  \bibinfo{person}{Kai Li}, {and} \bibinfo{person}{Li Fei-Fei}.}
  \bibinfo{year}{2009}\natexlab{}.
\newblock \showarticletitle{Imagenet: A large-scale hierarchical image
  database}. In \bibinfo{booktitle}{\emph{2009 IEEE conference on computer
  vision and pattern recognition}}. Ieee, \bibinfo{pages}{248--255}.
\newblock


\bibitem[\protect\citeauthoryear{FaCoY}{FaCoY}{2017}]%
        {facoy}
\bibfield{author}{\bibinfo{person}{FaCoY}.} \bibinfo{year}{2017}\natexlab{}.
\newblock \bibinfo{title}{https://github.com/facoy/facoy}.
\newblock
\newblock


\bibitem[\protect\citeauthoryear{Haykin}{Haykin}{1994}]%
        {haykin1994neural}
\bibfield{author}{\bibinfo{person}{Simon Haykin}.}
  \bibinfo{year}{1994}\natexlab{}.
\newblock \bibinfo{booktitle}{\emph{Neural networks: a comprehensive
  foundation}}.
\newblock \bibinfo{publisher}{Prentice Hall PTR}.
\newblock


\bibitem[\protect\citeauthoryear{He, Zhang, Ren, and Sun}{He
  et~al\mbox{.}}{2016}]%
        {he2016deep}
\bibfield{author}{\bibinfo{person}{Kaiming He}, \bibinfo{person}{Xiangyu
  Zhang}, \bibinfo{person}{Shaoqing Ren}, {and} \bibinfo{person}{Jian Sun}.}
  \bibinfo{year}{2016}\natexlab{}.
\newblock \showarticletitle{Deep residual learning for image recognition}. In
  \bibinfo{booktitle}{\emph{Proceedings of the IEEE conference on computer
  vision and pattern recognition}}. \bibinfo{pages}{770--778}.
\newblock


\bibitem[\protect\citeauthoryear{Jiang, Misherghi, Su, and Glondu}{Jiang
  et~al\mbox{.}}{2007}]%
        {jiang2007deckard}
\bibfield{author}{\bibinfo{person}{Lingxiao Jiang}, \bibinfo{person}{Ghassan
  Misherghi}, \bibinfo{person}{Zhendong Su}, {and} \bibinfo{person}{Stephane
  Glondu}.} \bibinfo{year}{2007}\natexlab{}.
\newblock \showarticletitle{Deckard: Scalable and accurate tree-based detection
  of code clones}. In \bibinfo{booktitle}{\emph{Proceedings of the 29th
  international conference on Software Engineering}}. IEEE Computer Society,
  \bibinfo{pages}{96--105}.
\newblock


\bibitem[\protect\citeauthoryear{Jiang and Su}{Jiang and Su}{2009}]%
        {jiang2009automatic}
\bibfield{author}{\bibinfo{person}{Lingxiao Jiang} {and}
  \bibinfo{person}{Zhendong Su}.} \bibinfo{year}{2009}\natexlab{}.
\newblock \showarticletitle{Automatic mining of functionally equivalent code
  fragments via random testing}. In \bibinfo{booktitle}{\emph{Proceedings of
  the eighteenth international symposium on Software testing and analysis}}.
  ACM, \bibinfo{pages}{81--92}.
\newblock


\bibitem[\protect\citeauthoryear{Juergens, Deissenboeck, and Hummel}{Juergens
  et~al\mbox{.}}{2010}]%
        {juergens2010code}
\bibfield{author}{\bibinfo{person}{Elmar Juergens}, \bibinfo{person}{Florian
  Deissenboeck}, {and} \bibinfo{person}{Benjamin Hummel}.}
  \bibinfo{year}{2010}\natexlab{}.
\newblock \showarticletitle{Code similarities beyond copy \& paste}. In
  \bibinfo{booktitle}{\emph{Software Maintenance and Reengineering (CSMR), 2010
  14th European Conference on}}. IEEE, \bibinfo{pages}{78--87}.
\newblock


\bibitem[\protect\citeauthoryear{Kamavisdar, Saluja, and Agrawal}{Kamavisdar
  et~al\mbox{.}}{2013}]%
        {kamavisdar2013survey}
\bibfield{author}{\bibinfo{person}{Pooja Kamavisdar}, \bibinfo{person}{Sonam
  Saluja}, {and} \bibinfo{person}{Sonu Agrawal}.}
  \bibinfo{year}{2013}\natexlab{}.
\newblock \showarticletitle{A survey on image classification approaches and
  techniques}.
\newblock \bibinfo{journal}{\emph{International Journal of Advanced Research in
  Computer and Communication Engineering}} \bibinfo{volume}{2},
  \bibinfo{number}{1} (\bibinfo{year}{2013}), \bibinfo{pages}{1005--1009}.
\newblock


\bibitem[\protect\citeauthoryear{Kamiya, Kusumoto, and Inoue}{Kamiya
  et~al\mbox{.}}{2002}]%
        {kamiya2002ccfinder}
\bibfield{author}{\bibinfo{person}{Toshihiro Kamiya}, \bibinfo{person}{Shinji
  Kusumoto}, {and} \bibinfo{person}{Katsuro Inoue}.}
  \bibinfo{year}{2002}\natexlab{}.
\newblock \showarticletitle{CCFinder: a multilinguistic token-based code clone
  detection system for large scale source code}.
\newblock \bibinfo{journal}{\emph{IEEE Transactions on Software Engineering}}
  \bibinfo{volume}{28}, \bibinfo{number}{7} (\bibinfo{year}{2002}),
  \bibinfo{pages}{654--670}.
\newblock


\bibitem[\protect\citeauthoryear{Kim, Jung, Kim, and Yi}{Kim
  et~al\mbox{.}}{2011}]%
        {kim-icse2011}
\bibfield{author}{\bibinfo{person}{H. Kim}, \bibinfo{person}{Y. Jung},
  \bibinfo{person}{S. Kim}, {and} \bibinfo{person}{K. Yi}.}
  \bibinfo{year}{2011}\natexlab{}.
\newblock \showarticletitle{{MeCC}: Memory comparison-based clone detector}. In
  \bibinfo{booktitle}{\emph{Proceedings of the 33rd International Conference on
  Software Engineering}}. \bibinfo{publisher}{IEEE}, \bibinfo{pages}{301--310}.
\newblock


\bibitem[\protect\citeauthoryear{Krinke}{Krinke}{2001}]%
        {krinke_identifying_2001}
\bibfield{author}{\bibinfo{person}{J. Krinke}.}
  \bibinfo{year}{2001}\natexlab{}.
\newblock \showarticletitle{Identifying similar code with program dependence
  graphs}. In \bibinfo{booktitle}{\emph{Proceedings {Eighth} {Working}
  {Conference} on {Reverse} {Engineering}}}. \bibinfo{pages}{301--309}.
\newblock


\bibitem[\protect\citeauthoryear{Krutz and Shihab}{Krutz and Shihab}{2013}]%
        {krutz_cccd:_2013}
\bibfield{author}{\bibinfo{person}{D.~E. Krutz} {and} \bibinfo{person}{E.
  Shihab}.} \bibinfo{year}{2013}\natexlab{}.
\newblock \showarticletitle{{CCCD}: {Concolic} code clone detection}. In
  \bibinfo{booktitle}{\emph{2013 20th {Working} {Conference} on {Reverse}
  {Engineering} ({WCRE})}}. \bibinfo{pages}{489--490}.
\newblock


\bibitem[\protect\citeauthoryear{Li, Feng, Zhuang, Meng, and Ryder}{Li
  et~al\mbox{.}}{2017}]%
        {li2017cclearner}
\bibfield{author}{\bibinfo{person}{Liuqing Li}, \bibinfo{person}{He Feng},
  \bibinfo{person}{Wenjie Zhuang}, \bibinfo{person}{Na Meng}, {and}
  \bibinfo{person}{Barbara Ryder}.} \bibinfo{year}{2017}\natexlab{}.
\newblock \showarticletitle{Cclearner: A deep learning-based clone detection
  approach}. In \bibinfo{booktitle}{\emph{2017 IEEE International Conference on
  Software Maintenance and Evolution (ICSME)}}. IEEE,
  \bibinfo{pages}{249--260}.
\newblock


\bibitem[\protect\citeauthoryear{Li, Xiao, Bassett, Xie, and Tillmann}{Li
  et~al\mbox{.}}{2016}]%
        {li_measuring_2016}
\bibfield{author}{\bibinfo{person}{Sihan Li}, \bibinfo{person}{Xusheng Xiao},
  \bibinfo{person}{Blake Bassett}, \bibinfo{person}{Tao Xie}, {and}
  \bibinfo{person}{Nikolai Tillmann}.} \bibinfo{year}{2016}\natexlab{}.
\newblock \showarticletitle{Measuring {Code} {Behavioral} {Similarity} for
  {Programming} and {Software} {Engineering} {Education}}. In
  \bibinfo{booktitle}{\emph{Proceedings of the 38th {International}
  {Conference} on {Software} {Engineering} {Companion}}}.
  \bibinfo{publisher}{ACM}, \bibinfo{address}{New York, NY, USA},
  \bibinfo{pages}{501--510}.
\newblock
\showISBNx{978-1-4503-4205-6}


\bibitem[\protect\citeauthoryear{Li, Lu, Myagmar, and Zhou}{Li
  et~al\mbox{.}}{2004}]%
        {li2004cp}
\bibfield{author}{\bibinfo{person}{Zhenmin Li}, \bibinfo{person}{Shan Lu},
  \bibinfo{person}{Suvda Myagmar}, {and} \bibinfo{person}{Yuanyuan Zhou}.}
  \bibinfo{year}{2004}\natexlab{}.
\newblock \showarticletitle{{CP}-{Miner}: a tool for finding copy-paste and
  related bugs in operating system code}. In
  \bibinfo{booktitle}{\emph{Proceedings of the 6th conference on {Symposium} on
  {Opearting} {Systems} {Design} \& {Implementation} - {Volume} 6}}.
  \bibinfo{publisher}{USENIX Association}, \bibinfo{address}{Berkeley, CA,
  USA}, \bibinfo{pages}{20--20}.
\newblock


\bibitem[\protect\citeauthoryear{Li, Zou, Xu, Ou, Jin, Wang, Deng, and
  Zhong}{Li et~al\mbox{.}}{2018}]%
        {li2018vuldeepecker}
\bibfield{author}{\bibinfo{person}{Zhen Li}, \bibinfo{person}{Deqing Zou},
  \bibinfo{person}{Shouhuai Xu}, \bibinfo{person}{Xinyu Ou},
  \bibinfo{person}{Hai Jin}, \bibinfo{person}{Sujuan Wang},
  \bibinfo{person}{Zhijun Deng}, {and} \bibinfo{person}{Yuyi Zhong}.}
  \bibinfo{year}{2018}\natexlab{}.
\newblock \showarticletitle{VulDeePecker: A deep learning-based system for
  vulnerability detection}.
\newblock \bibinfo{journal}{\emph{arXiv preprint arXiv:1801.01681}}
  (\bibinfo{year}{2018}).
\newblock


\bibitem[\protect\citeauthoryear{Liu, Chen, Han, and Yu}{Liu
  et~al\mbox{.}}{2006}]%
        {liu_gplag:_2006}
\bibfield{author}{\bibinfo{person}{Chao Liu}, \bibinfo{person}{Chen Chen},
  \bibinfo{person}{Jiawei Han}, {and} \bibinfo{person}{Philip~S. Yu}.}
  \bibinfo{year}{2006}\natexlab{}.
\newblock \showarticletitle{{GPLAG}: {Detection} of {Software} {Plagiarism} by
  {Program} {Dependence} {Graph} {Analysis}}. In
  \bibinfo{booktitle}{\emph{Proceedings of the 12th {ACM} {SIGKDD}
  {International} {Conference} on {Knowledge} {Discovery} and {Data}
  {Mining}}}. \bibinfo{publisher}{ACM}, \bibinfo{address}{New York, NY, USA},
  \bibinfo{pages}{872--881}.
\newblock
\showISBNx{978-1-59593-339-3}


\bibitem[\protect\citeauthoryear{Lu and Weng}{Lu and Weng}{2007}]%
        {lu2007survey}
\bibfield{author}{\bibinfo{person}{Dengsheng Lu} {and} \bibinfo{person}{Qihao
  Weng}.} \bibinfo{year}{2007}\natexlab{}.
\newblock \showarticletitle{A survey of image classification methods and
  techniques for improving classification performance}.
\newblock \bibinfo{journal}{\emph{International journal of Remote sensing}}
  \bibinfo{volume}{28}, \bibinfo{number}{5} (\bibinfo{year}{2007}),
  \bibinfo{pages}{823--870}.
\newblock


\bibitem[\protect\citeauthoryear{Marcus and Maletic}{Marcus and
  Maletic}{2001}]%
        {marcus2001identification}
\bibfield{author}{\bibinfo{person}{Andrian Marcus} {and}
  \bibinfo{person}{Jonathan~I Maletic}.} \bibinfo{year}{2001}\natexlab{}.
\newblock \showarticletitle{Identification of high-level concept clones in
  source code}. In \bibinfo{booktitle}{\emph{Proceedings 16th Annual
  International Conference on Automated Software Engineering (ASE 2001)}}.
  IEEE, \bibinfo{pages}{107--114}.
\newblock


\bibitem[\protect\citeauthoryear{Martin}{Martin}{2000}]%
        {martin2000design}
\bibfield{author}{\bibinfo{person}{Robert~C Martin}.}
  \bibinfo{year}{2000}\natexlab{}.
\newblock \showarticletitle{Design principles and design patterns}.
\newblock \bibinfo{journal}{\emph{Object Mentor}} \bibinfo{volume}{1},
  \bibinfo{number}{34} (\bibinfo{year}{2000}), \bibinfo{pages}{597}.
\newblock


\bibitem[\protect\citeauthoryear{Mikolov, Chen, Corrado, and Dean}{Mikolov
  et~al\mbox{.}}{2013a}]%
        {mikolov2013efficient}
\bibfield{author}{\bibinfo{person}{Tomas Mikolov}, \bibinfo{person}{Kai Chen},
  \bibinfo{person}{Greg Corrado}, {and} \bibinfo{person}{Jeffrey Dean}.}
  \bibinfo{year}{2013}\natexlab{a}.
\newblock \showarticletitle{Efficient estimation of word representations in
  vector space}.
\newblock \bibinfo{journal}{\emph{arXiv preprint arXiv:1301.3781}}
  (\bibinfo{year}{2013}).
\newblock


\bibitem[\protect\citeauthoryear{Mikolov, Sutskever, Chen, Corrado, and
  Dean}{Mikolov et~al\mbox{.}}{2013b}]%
        {mikolov2013distributed}
\bibfield{author}{\bibinfo{person}{Tomas Mikolov}, \bibinfo{person}{Ilya
  Sutskever}, \bibinfo{person}{Kai Chen}, \bibinfo{person}{Greg~S Corrado},
  {and} \bibinfo{person}{Jeff Dean}.} \bibinfo{year}{2013}\natexlab{b}.
\newblock \showarticletitle{Distributed representations of words and phrases
  and their compositionality}. In \bibinfo{booktitle}{\emph{Advances in neural
  information processing systems}}. \bibinfo{pages}{3111--3119}.
\newblock


\bibitem[\protect\citeauthoryear{Mou, Li, Zhang, Wang, and Jin}{Mou
  et~al\mbox{.}}{2016}]%
        {mou2016convolutional}
\bibfield{author}{\bibinfo{person}{Lili Mou}, \bibinfo{person}{Ge Li},
  \bibinfo{person}{Lu Zhang}, \bibinfo{person}{Tao Wang}, {and}
  \bibinfo{person}{Zhi Jin}.} \bibinfo{year}{2016}\natexlab{}.
\newblock \showarticletitle{Convolutional neural networks over tree structures
  for programming language processing}. In \bibinfo{booktitle}{\emph{Thirtieth
  AAAI Conference on Artificial Intelligence}}.
\newblock


\bibitem[\protect\citeauthoryear{Pan and Yang}{Pan and Yang}{2009}]%
        {pan2009survey}
\bibfield{author}{\bibinfo{person}{Sinno~Jialin Pan} {and}
  \bibinfo{person}{Qiang Yang}.} \bibinfo{year}{2009}\natexlab{}.
\newblock \showarticletitle{A survey on transfer learning}.
\newblock \bibinfo{journal}{\emph{IEEE Transactions on knowledge and data
  engineering}} \bibinfo{volume}{22}, \bibinfo{number}{10}
  (\bibinfo{year}{2009}), \bibinfo{pages}{1345--1359}.
\newblock


\bibitem[\protect\citeauthoryear{Paszke, Gross, Chintala, and Chanan}{Paszke
  et~al\mbox{.}}{2017}]%
        {paszke2017pytorch}
\bibfield{author}{\bibinfo{person}{Adam Paszke}, \bibinfo{person}{Sam Gross},
  \bibinfo{person}{Soumith Chintala}, {and} \bibinfo{person}{Gregory Chanan}.}
  \bibinfo{year}{2017}\natexlab{}.
\newblock \showarticletitle{Pytorch}.
\newblock \bibinfo{journal}{\emph{Computer software. Vers. 0.3}}
  \bibinfo{volume}{1} (\bibinfo{year}{2017}).
\newblock


\bibitem[\protect\citeauthoryear{Ragkhitwetsagul, Krinke, and
  Marnette}{Ragkhitwetsagul et~al\mbox{.}}{2018}]%
        {ragkhitwetsagul2018picture}
\bibfield{author}{\bibinfo{person}{Chaiyong Ragkhitwetsagul},
  \bibinfo{person}{Jens Krinke}, {and} \bibinfo{person}{Bruno Marnette}.}
  \bibinfo{year}{2018}\natexlab{}.
\newblock \showarticletitle{A picture is worth a thousand words: Code clone
  detection based on image similarity}. In \bibinfo{booktitle}{\emph{2018 IEEE
  12th International Workshop on Software Clones (IWSC)}}. IEEE,
  \bibinfo{pages}{44--50}.
\newblock


\bibitem[\protect\citeauthoryear{Roy, Cordy, and Koschke}{Roy
  et~al\mbox{.}}{2009}]%
        {roy2009comparison}
\bibfield{author}{\bibinfo{person}{Chanchal~K Roy}, \bibinfo{person}{James~R
  Cordy}, {and} \bibinfo{person}{Rainer Koschke}.}
  \bibinfo{year}{2009}\natexlab{}.
\newblock \showarticletitle{Comparison and evaluation of code clone detection
  techniques and tools: A qualitative approach}.
\newblock \bibinfo{journal}{\emph{Science of Computer Programming}}
  \bibinfo{volume}{74}, \bibinfo{number}{7} (\bibinfo{year}{2009}),
  \bibinfo{pages}{470--495}.
\newblock


\bibitem[\protect\citeauthoryear{Saini, Farmahinifarahani, Lu, Baldi, and
  Lopes}{Saini et~al\mbox{.}}{2018}]%
        {saini2018oreo}
\bibfield{author}{\bibinfo{person}{Vaibhav Saini}, \bibinfo{person}{Farima
  Farmahinifarahani}, \bibinfo{person}{Yadong Lu}, \bibinfo{person}{Pierre
  Baldi}, {and} \bibinfo{person}{Cristina~V Lopes}.}
  \bibinfo{year}{2018}\natexlab{}.
\newblock \showarticletitle{Oreo: Detection of clones in the twilight zone}. In
  \bibinfo{booktitle}{\emph{Proceedings of the 2018 26th ACM Joint Meeting on
  European Software Engineering Conference and Symposium on the Foundations of
  Software Engineering}}. ACM, \bibinfo{pages}{354--365}.
\newblock


\bibitem[\protect\citeauthoryear{Su, Bell, Harvey, Sethumadhavan, Kaiser, and
  Jebara}{Su et~al\mbox{.}}{2016a}]%
        {su2016code}
\bibfield{author}{\bibinfo{person}{Fang-Hsiang Su}, \bibinfo{person}{Jonathan
  Bell}, \bibinfo{person}{Kenneth Harvey}, \bibinfo{person}{Simha
  Sethumadhavan}, \bibinfo{person}{Gail Kaiser}, {and} \bibinfo{person}{Tony
  Jebara}.} \bibinfo{year}{2016}\natexlab{a}.
\newblock \showarticletitle{Code relatives: detecting similarly behaving
  software}. In \bibinfo{booktitle}{\emph{Proceedings of the 2016 24th ACM
  SIGSOFT International Symposium on Foundations of Software Engineering}}.
  ACM, \bibinfo{pages}{702--714}.
\newblock


\bibitem[\protect\citeauthoryear{Su, Bell, Harvey, Sethumadhavan, Kaiser, and
  Jebara}{Su et~al\mbox{.}}{2016b}]%
        {Su:2016:CodeRelatives}
\bibfield{author}{\bibinfo{person}{Fang-Hsiang Su}, \bibinfo{person}{Jonathan
  Bell}, \bibinfo{person}{Kenneth Harvey}, \bibinfo{person}{Simha
  Sethumadhavan}, \bibinfo{person}{Gail Kaiser}, {and} \bibinfo{person}{Tony
  Jebara}.} \bibinfo{year}{2016}\natexlab{b}.
\newblock \showarticletitle{Code Relatives: Detecting Similarly Behaving
  Software}. In \bibinfo{booktitle}{\emph{Proceedings of the 2016 24th ACM
  SIGSOFT International Symposium on Foundations of Software Engineering}}
  \emph{(\bibinfo{series}{FSE 2016})}. \bibinfo{publisher}{ACM},
  \bibinfo{pages}{702--714}.
\newblock
\showISBNx{978-1-4503-4218-6}


\bibitem[\protect\citeauthoryear{Svajlenko, Islam, Keivanloo, Roy, and
  Mia}{Svajlenko et~al\mbox{.}}{2014}]%
        {svajlenko2014towards}
\bibfield{author}{\bibinfo{person}{Jeffrey Svajlenko},
  \bibinfo{person}{Judith~F Islam}, \bibinfo{person}{Iman Keivanloo},
  \bibinfo{person}{Chanchal~K Roy}, {and} \bibinfo{person}{Mohammad~Mamun
  Mia}.} \bibinfo{year}{2014}\natexlab{}.
\newblock \showarticletitle{Towards a big data curated benchmark of
  inter-project code clones}. In \bibinfo{booktitle}{\emph{Software Maintenance
  and Evolution (ICSME), 2014 IEEE International Conference on}}. IEEE,
  \bibinfo{pages}{476--480}.
\newblock


\bibitem[\protect\citeauthoryear{Tufano, Watson, Bavota, Di~Penta, White, and
  Poshyvanyk}{Tufano et~al\mbox{.}}{2018}]%
        {tufano2018deep}
\bibfield{author}{\bibinfo{person}{Michele Tufano}, \bibinfo{person}{Cody
  Watson}, \bibinfo{person}{Gabriele Bavota}, \bibinfo{person}{Massimiliano
  Di~Penta}, \bibinfo{person}{Martin White}, {and} \bibinfo{person}{Denys
  Poshyvanyk}.} \bibinfo{year}{2018}\natexlab{}.
\newblock \showarticletitle{Deep learning similarities from different
  representations of source code}. In \bibinfo{booktitle}{\emph{2018 IEEE/ACM
  15th International Conference on Mining Software Repositories (MSR)}}. IEEE,
  \bibinfo{pages}{542--553}.
\newblock


\bibitem[\protect\citeauthoryear{Wei and Li}{Wei and Li}{2017}]%
        {wei2017supervised}
\bibfield{author}{\bibinfo{person}{Huihui Wei} {and} \bibinfo{person}{Ming
  Li}.} \bibinfo{year}{2017}\natexlab{}.
\newblock \showarticletitle{Supervised Deep Features for Software Functional
  Clone Detection by Exploiting Lexical and Syntactical Information in Source
  Code.}. In \bibinfo{booktitle}{\emph{IJCAI}}. \bibinfo{pages}{3034--3040}.
\newblock


\bibitem[\protect\citeauthoryear{Yi~Gao and Cai.}{Yi~Gao and Cai.}{2019}]%
        {gao2019teccd}
\bibfield{author}{\bibinfo{person}{Shuang Liu Lin Yang Sang~Wei Yi~Gao,
  Zan~Wang} {and} \bibinfo{person}{Yuanfang Cai.}}
  \bibinfo{year}{2019}\natexlab{}.
\newblock \showarticletitle{TECCD: A Tree Embedding Approach for Code Clone
  Detection}. In \bibinfo{booktitle}{\emph{2019 IEEE International Conference
  on Software Maintenance and Evolution (ICSME)}}. IEEE.
\newblock


\bibitem[\protect\citeauthoryear{Zhang, Wang, Zhang, Sun, Wang, and Liu}{Zhang
  et~al\mbox{.}}{2019}]%
        {zhang2019novel}
\bibfield{author}{\bibinfo{person}{Jian Zhang}, \bibinfo{person}{Xu Wang},
  \bibinfo{person}{Hongyu Zhang}, \bibinfo{person}{Hailong Sun},
  \bibinfo{person}{Kaixuan Wang}, {and} \bibinfo{person}{Xudong Liu}.}
  \bibinfo{year}{2019}\natexlab{}.
\newblock \showarticletitle{A novel neural source code representation based on
  abstract syntax tree}. In \bibinfo{booktitle}{\emph{Proceedings of the 41st
  International Conference on Software Engineering}}. IEEE Press,
  \bibinfo{pages}{783--794}.
\newblock


\bibitem[\protect\citeauthoryear{Zhu, Sobihani, and Guo}{Zhu
  et~al\mbox{.}}{2015}]%
        {zhu2015long}
\bibfield{author}{\bibinfo{person}{Xiaodan Zhu}, \bibinfo{person}{Parinaz
  Sobihani}, {and} \bibinfo{person}{Hongyu Guo}.}
  \bibinfo{year}{2015}\natexlab{}.
\newblock \showarticletitle{Long short-term memory over recursive structures}.
  In \bibinfo{booktitle}{\emph{International Conference on Machine Learning}}.
  \bibinfo{pages}{1604--1612}.
\newblock


\end{thebibliography}
\appendix





\end{document}